\documentclass[fleqn,usenatbib]{mnras}

\usepackage[T1]{fontenc}

\DeclareRobustCommand{\VAN}[3]{#2}
\let\VANthebibliography\thebibliography
\def\thebibliography{\DeclareRobustCommand{\VAN}[3]{##3}\VANthebibliography}

\usepackage{graphicx}	
\usepackage{amsmath}	
\usepackage{amssymb}	
\usepackage{float}
\usepackage{longtable}
\usepackage{natbib}
\usepackage[toc,page]{appendix}
\usepackage{subfig}
\usepackage{multicol}
\usepackage{multirow}
\usepackage{xcolor}
\usepackage{lscape}

\title[The search for gas in debris discs]{The search for gas in debris discs: ALMA detection of CO gas in HD 36546}

\author[I. Rebollido et al.]{
Isabel Rebollido,$^{1}$\thanks{E-mail: irebollido@stsci.edu} 
\'Alvaro Ribas$^{2}$, 
Itziar de Gregorio-Monsalvo$^{2}$, 
Eva Villaver$^{3}$, 
Benjam\'in Montesinos$^{3}$, 
\newauthor
Christine Chen$^{1}$, 
H\'ector Canovas$^{4}$, 
Thomas Henning$^{5}$, 
Attila Mo\'or$^{6}$, 
Marshall Perrin$^{1}$, 
\newauthor
Pablo Rivi\`ere-Marichalar$^{7}$,
and Carlos Eiroa$^{8}$
\\
$^{1}$Space Telescope Science Institute, 3700 San Martin Drive, Baltimore, MD 21218, USA\\
$^{2}$European Southern Observatory (ESO), Alonso de C\'ordova 3107, Vitacura, Casilla 19001,Santiago de Chile,Chile\\
$^{3}$ Centro de Astrobiolog\'ia (CAB, CSIC-INTA), ESAC Campus Camino Bajo del Castillo, s/n, Villanueva de la Cañada,
28692 Madrid, Spain\\
$^{4}$ Telespazio UK for the European Space Agency (ESA), European Space Astronomy Centre
(ESAC), \\Camino Bajo del Castillo s/n, 28692 Villanueva de la Ca\~nada, Madrid, Spain\\
$^{5}$ Max-Planck-Institut f\"ur Astronomie (MPIA), K\"onigstuhl 17, 69117 Heidelberg, Germany\\
$^{6}$ Konkoly Observatory, Research Centre for Astronomy and Earth Sciences, E\"otv\"os Lor\'and Research Network (ELKH), \\Konkoly-Thege Mikl\'os \'ut 15-17, H-1121 Budapest, Hungary\\
$^{7}$ Observatorio Astronómico Nacional (OAN-IGN)-Observatorio de Madrid, Alfonso XII, 3, 28014 Madrid, Spain\\
$^{8}$ Private Researcher (formerly at Universidad Aut\'onoma de Madrid)
} 

\date{Accepted XXX. Received YYY; in original form ZZZ}

\pubyear{2021}

\usepackage{newtxtext,newtxmath}

\begin{document}
\label{firstpage}
\pagerange{\pageref{firstpage}--\pageref{lastpage}}
\maketitle

\begin{abstract}
Debris discs represent the last stages of planet formation and as such are expected to be depleted of primordial gas. Nonetheless, in the last few years the presence of cold gas has been reported in $\sim$ 20 debris discs from far-IR to (sub-)mm observations and hot gas has been observed in the optical spectra of debris discs for decades. While the origin of this gas is still uncertain, most evidences point towards a secondary origin, as a result of collisions and evaporation of small bodies in the disc. In this paper, we present ALMA observations aimed at the detection of CO gas in a sample of 8 debris discs with optical gas detections. We report the detection of CO ($^{12}$CO and $^{13}$CO) gas in HD 36546, the brightest and youngest disc in our sample, and provide upper limits to the presence of gas in the remaining seven discs.

\end{abstract}




\section{Introduction}

Classically, the environment of main-sequence stars has been assumed to be essentially gas free, as it is expected that primordial gas is depleted, either by photoevaporation, formation of planets, or accreted by the star at the end of the protoplanetary phase \citep[see e.g.][for a review in debris discs]{wyatt18,hughes18}. However, we have known for almost four decades that there is hot gas in the inner regions (at few R$_*$) of some stars detected as superimposed narrow stable absorptions in certain metallic lines \citep{hobbs85,hobbs88}. In some cases, these narrow features are accompanied by transient absorptions of different equivalent widths, both blue- and red-shifted with respect to the stellar radial velocity. The most famous example is $\beta$ Pictoris, where the detection of narrow variable features was interpreted at first as \textit{falling evaporating bodies} (or FEBs) and later as exocomets \citep[see e.g. ][]{ferlet87,kiefer14b}.
Both the stable gas and the presence of exocomet signatures is often linked to the presence of a debris disc in the system, with excess at long wavelengths \citep[see e.g. ][]{Rebollido20}. These debris discs can in most cases be characterised with two temperature belts, at different distances from the star, indicating an architecture similar to the solar system \citep{chen14}. More recently,  thanks to the availability  of far-IR and (sub-)mm facilities like the \emph{Herschel Space Observatory} \citep{pilbratt10}, the \emph{Atacama Pathfinder EXperiment} (APEX) or the \emph{Atacama Large Millimeter/submillimeter Array} (ALMA), cold gas has also been detected in the environment of debris discs stars, located at several tens au from the central object \citep[e.g.][]{zuckerman95,hughes08,moor11b,roberge13,riviere14,moor15a,liemansifry16,marino16,moor17,matra17b,kral19}. 

The origin of both cold and hot gas is still debated, specially in the discs with larger gas content. The most likely explanation is a secondary origin \citep[e.g.][]{marino16,matra17a,kral19} as a result of outgassing or collisions between minor bodies. However, in some young systems with a particularly high CO content, it has also been suggested that while the dust is of secondary origin, the gas 
is composed predominantly of material leftover from the primordial disc \citep{kospal13}.
Furthermore, \cite{Rebollido18} 
detected hot gas in absorption in cold-gas-bearing debris discs with edge-on orientation, but not in those with face-on orientation. This was attributed to a geometrical observational effect, implying hot and cold gas are simultaneously present in discs despite the detectability, and possibly have a common secondary origin where small bodies in the system would be releasing gas (via evaporation or collisions) at different locations in the disc. The presence of exocomets in some of these objects strengthens the theory that small objects and planetesimals could play a key role in the replenishment of secondary gas.

In this work, we present the results of an ALMA survey to search for cold CO gas in a sample of debris disc stars with hot gas detected through optical spectroscopy. The goal is to test the hypothesis of the simultaneous presence and maybe a common origin for both types of gas.
We report the detection of CO in one of the targets, HD 36546, the youngest in our sample.

 \
\section{Sample and observations}
\subsection{Sample} 
\label{sect:sample}
Our target selection required the stars to have debris discs, evidence of hot circumstellar gas \citep[either variable or stable, see][]{Rebollido20}, to be accessible from the ALMA site, and to not have been observed before for a cold-gas detection. Our final list is composed of eight stars hosting debris discs with IR luminosities on average one order of magnitude lower than the known cold-gas bearing debris discs and with SEDs compatible with a two-belt dust structure, indicative of a system architecture comparable to the Solar System asteroid and Kuiper belts. These features are similar to those present in the objects with hot and cold gas already confirmed \citep{Rebollido18}. Furthermore, six of the sources show variability in spectroscopic observations, attributable to the presence of exocomets. Table \ref{tab:sample} lists the objects observed and some properties, including age, distance and fractional luminosity. Spectral energy distributions of the objects in the sample are shown in Fig. \ref{fig:fit} and \ref{fig:seds}. 

\begin{table*}
    \centering
    \begin{tabular}{l l l c c c c c c c}
\hline
Name & RA~(J2000) & DEC~(J2000) & Sp. type & Distance($^{*}$)  & V & Age & L$_{\rm IR}$/L$_{*}$ & F$_{\nu}$(1.3 mm) & $^{12}$CO (2-1)\\
\hline
&hh:mm:ss & dd:mm:ss & &pc & (mag) &Myr & & mJy & mJy~km~s$^{-1}$  \\
\hline
\hline
HD 5267   & 00:54:35.23 & +19:11:18.3  & A1V               & 76.8  $\pm$ 4.4 & 5.79 & 200 (1) & 3.9$\cdot$10$^{-5}$(1)& <0.11 & <9.5 \\
HD 36546  & 05:33:30.76 & +24:37:43.72 & B8V$^{(\dagger)}$ & 100.2 $\pm$ 0.4 & 6.95 & 3-10 (2)  & 3.4$\cdot$10$^{-3}$(2)& 2.59$\pm$0.05 & (2.67 $\pm$ 0.04)$\times$10$^{3}$ \\
HD 37306  & 05:37:08.77 & -11:46:31.9  & A2V               & 69.6  $\pm$ 0.2 & 6.09 & 38-48 (3)  & 1.2$\cdot$10$^{-4}$(1) & <0.11 & <9.5 \\ 
HD 110411 & 12:41:53.06 & +10:14:08.3  & A3V               & 38.9  $\pm$ 0.2 & 4.88 & 86 (1)  & 6.4$\cdot$10$^{-5}$(3) & 0.29$\pm$0.06 & <10.5\\
HD 145964 & 16:14:28.88 & -21:06:27.5  & B9V               & 113.0 $\pm$ 0.6 & 6.41 & 11 (1)  & 1.5$\cdot$10$^{-5}$(1) & <0.13 & <13.5\\
HD 158352 & 17:28:49.66 & +00:19:50.3  & A8Vp              & 63.8  $\pm$ 0.3 & 5.41 & 890 (1) & 9.3$\cdot$10$^{-5}$(4) & 1.63$\pm$0.15 & <14.0\\
HD 182919 & 19:26:13.25 & +20:05:51.8  & A0V               & 72.1  $\pm$ 0.2 & 5.59 & 198 (1) & 3.4$\cdot$10$^{-5}$(1) & <0.08 & <10.5\\
HD 183324 & 19:29:00.99 & +01:57:01.6  & A0IV              & 60.4  $\pm$ 0.2 & 5.79 & 140 (1) & 1.8$\cdot$10$^{-5}$(5) & <0.13 & <14.0\\
\hline
    \end{tabular}
\caption{
    Objects in the observed sample. The 1.3\,mm and $^{12}$CO fluxes include the 5 $\sigma$ upper limits and detections of the integrated intensity over the whole emission area. Numbers in parenthesis denote the corresponding references. \\
    References. Age: (1) \protect\cite{chen14}; (2) \protect\cite{lisse17}; (3) \protect\cite{bell15}. L$_{\rm IR}$/L$_{*}$: (1) \protect\cite{chen14}; (2) \protect\cite{lisse17}; (3) \protect\cite{thureau14}; (4) \protect\cite{roberge08}; (5) \protect\cite{draper16}\\
    $^{(\dagger)}$ This star has also been reported in the literature as an A1 star \protect\citep{lisse17}.\\
    $^{*}$ Distances are calculated from Gaia DR2-EDR3 parallaxes \protect\cite{GAIADR2,GaiaDR3}.
    }
\label{tab:sample}
\end{table*}

\subsection{Observations and data reduction}

The sample was observed with ALMA during Cycle 7 (project 2019.1.01517.S, PI: I. Rebollido) between November 25$^{\rm{th}}$ and December 22$^{\rm{nd}}$, 2019. Observations were performed in Band~6, aiming to detect both continuum and CO J=2-1 emission from the discs: the correlator setup included six spectral windows, three of them centered at the $^{12}$CO, $^{13}$CO, and C$^{18}$O J=2-1 transitions with a bandwidth of 468.75\,MHz and a channel width of 244\,kHz, and three additional spectral windows to probe the continuum (two of them centered at 217 and 234\,GHz with a 2\,GHz bandwidth, and one more located at 231.4\,GHz with a bandwidth of 468.75\,MHz). We used the most compact configuration of the main array (C43-1, with baselines ranging from 15 to 300\,m) to maximize sensitivity, yielding beam sizes $\sim$1.5\arcsec x1.0\arcsec.

The calibrated observations provided by the observatory were used to image the continuum using the \texttt{tclean} task of the CASA software \citep{McMullin07} version 5.4. We produced images with map sizes of $300\times300$ pixels and a pixel size of 0.1\,\arcsec using Briggs weighting, and explored different values of the \texttt{robust} parameter. Resolved continuum detections at the nominal source positions were found for HD~36546, HD~158352 and HD~110411, and offset emission was also detected for HD~37306 (multiple sources) and HD~182919, possibly associated with (sub-)mm background objects. The final 1.33\,mm continuum images (see Fig. \ref{fig:HD36546} and Fig. \ref{fig:sample}) were produced with \texttt{robust}=0.5 as a compromise between sensitivity and resolution, except for the case of HD 110411 and HD 158352, where images were produced with \texttt{robust}=2 in order to improve the signal to noise ratio. In the case of HD~36546, we also performed one single round of phase-only self-calibration on each of its two individual observations  by combining all spectral windows (after masking channels with potential line emission) and scans, which increased its peak S/N from 47 to 65. For HD~36546 and HD~158352, image cubes of the three targeted CO lines were produced using the continuum-subtracted visibilities around the corresponding lines. For the rest of the objects in the sample, cubes of the $^{12}$CO line were also produced in order to estimate upper limits.

We retrieved the peak and integrated flux, the angular size and the inclination using the fitting function in the CASA viewer for the continuum images and the zeroth moment maps of the $^{12}$CO and $^{13}$CO lines.

\section{Results}
\subsection{General results}

Out of the eight objects in the sample, only HD 36546 shows evidence of CO gas in the disc and was also detected in continuum. The remaining seven objects show no evidence of CO gas, and three of them have also no detection in the continuum (see Table \ref{tab:sample} and Fig. \ref{fig:sample}). There is, however, detection of continuum emission in the field of view of HD 37306, HD 110411, HD 158352 and HD 182919.

When there is no CO detection Table \ref{tab:sample} includes upper limits (5 $\sigma$) for both the $^{12}$CO and continuum measurements. 
Three point sources are detected around 10 $\arcsec$ north of the nominal position of HD 37306 in continuum emission (see Fig. \ref{fig:sample}), two of them with similar fluxes of 0.6$\pm$0.1 mJy; and a third one with 0.17$\pm$0.05 mJy. They do not seem to be related to the source, and are most likely background objects. No sources were found with similar coordinates in the Gaia \footnote{https://gea.esac.esa.int/archive/} or NASA/IPAC Extragalactic Database archives \footnote{https://ned.ipac.caltech.edu}.  

There is also a point source detection in the continuum observations at $\sim$ 10$\arcsec$ from the nominal position of HD 182919, that is not compatible with emission from the target star. A search in the Gaia archive has revealed one object compatible with the position of the detected emission (Gaia EDR3 4515892996340398336) with a faint $G$ magnitude ($\sim$ 20), and therefore the emission is likely coming from this background object. 

There are three sources in our sample for which there are associated continuum detections: HD 36546, HD 110411 and HD 158352. HD 36546 continuum emission is detailed in the next subsection.
HD 158352 shows an elongated, low SNR ($\sim$ 8), continuum emission that could correspond to the thermal emission of a dusty circumstellar disc. 
The integrated flux averaged over the whole emission area is given in Table \ref{tab:sample}. After gaussian deconvolution, the size of the major axis of the disc is 540 $\pm$ 70 au. Taking the temperature for the dust given in \cite{roberge08}, 76.1 K, it yields a dust mass of 5$\cdot$ 10$^{-2}$ M$_{\earth}$ (see procedure in Sect. \ref{sect:dust}).

HD 110411 shows a tentative detection of $\sim$ 5 $\sigma$ at the position of the star. The flux density integrated over all the disc structure is shown in Table \ref{tab:sample}, but further observations are needed to confirm the presence of mm-sized dust in the system. If we consider the obtained flux and the temperature given for the cold component in \cite{thureau14}, 68 K, we obtain a dust mass for the system of 3$\cdot$ 10$^{-3}$ M$_{\earth}$.

\subsection{HD 36546}

\begin{table*}
    \centering
    \begin{tabular}{l c c c c c}

\hline
  & Peak Intensity$^*$ & Flux$^*$ & a &  Incl. & M\\

 &  &     & (au)  & (deg) & (M$_{\earth}$)\\
\hline
Cont. & 1.33 $\pm$ 0.03 mJy & 2.59$\pm$0.05 mJy & 187.7 $\pm$ 6.2 & 79.0 $\pm$ 1.5 & (9.0$\pm$1.0)$\cdot$10$^{-2}$\\
$^{12}$CO & 1.42 $\pm$ 0.02 Jy km s$^{-1}$ beam$^{-1}$ & 2.67 $\pm$ 0.04 Jy km s$^{-1}$ & 216 $\pm$ 4 & 78.3 $\pm$ 1.2 &\multirow{2}{*}{(3.2$\pm$1.2)$\cdot$10$^{-3}$} \\  
$^{13}$CO & 0.07 $\pm$ 0.01 Jy km s$^{-1}$ beam$^{-1}$& 0.15 $\pm$ 0.02 Jy km s$^{-1}$ & 259 $\pm$ 84& 75.1 $\pm$ 6.2 & \\

\hline
    \end{tabular}
    \caption{Continuum and gas detections for HD 36546. Table shows the peak intensities,  fluxes, major axis (a), inclination and mass. The $^{13}$CO observations were partially unresolved, and therefore no inclination data is provided. \\
    $^*$ Uncertainties do not account for the absolute flux calibration uncertainty ($\leq$ 10\% in this band).}
    \label{tab:hd36546}
\end{table*}

\subsubsection{Continuum}
\label{sect:dust}
\begin{figure}
    \centering
	\includegraphics[width=\hsize]{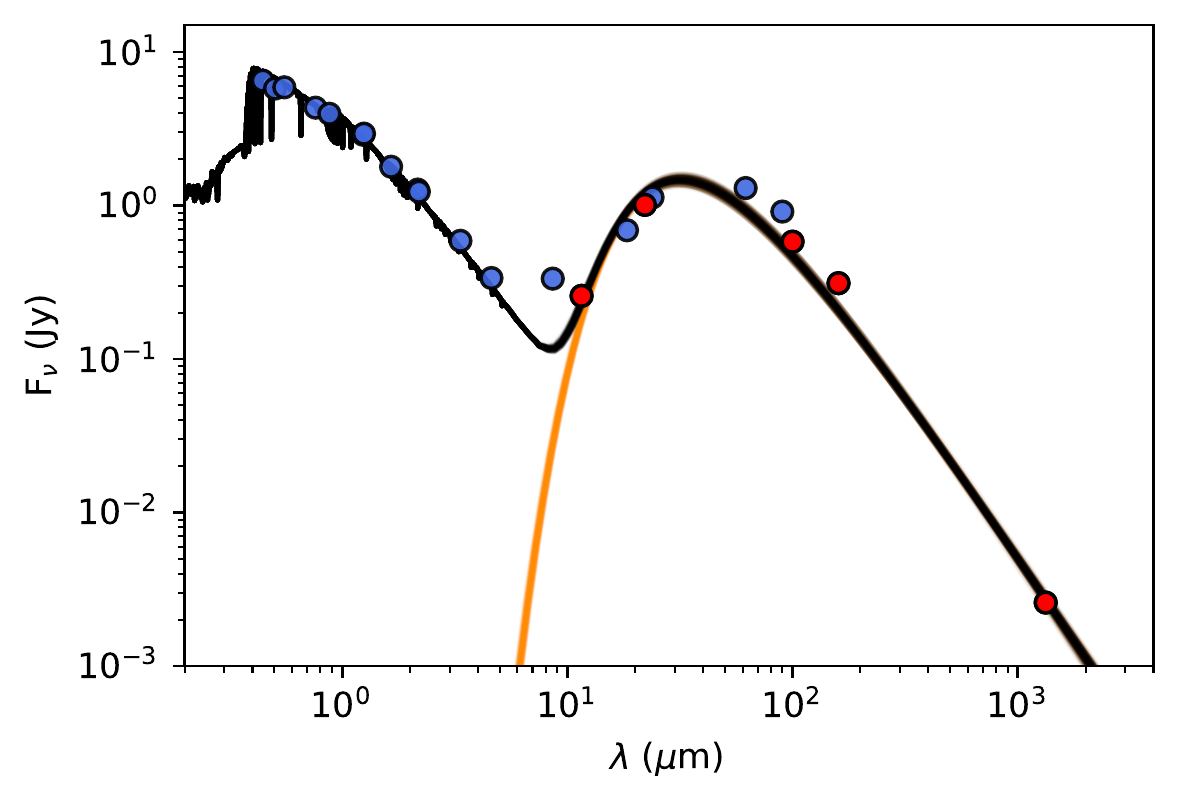}
    \caption{Modified black body fit of the observed photometry of HD 36546. Available photometry is shown as blue dots, and those used to fit the modified black body are marked in red. The orange lines are 100 modified black bodies randomly drawn from the posterior distribution, and are added to the corresponding Kurucz photosphere to create the full model (black lines). The adopted stellar parameters come from \protect\cite{Rebollido20}.
    }
    \label{fig:fit}
\end{figure}

We detect 1.33 mm continuum extended emission originating from HD 36546 (see Fig. \ref{fig:HD36546}). Peak and integrated flux, size, and inclination are reported in Table \ref{tab:hd36546}. Once deconvolved from the beam, the major axis of the disc spans 180 au, compatible with the size reported from scattered light observations \citep[semi-major axis of 85 au, ][]{currie17}. The inclination is directly provided by the CASA tool, as obtained from the ratio of the major and minor axis deconvolved from the beam.
The total mass can be estimated by assuming an optically thin dust disc as $M_d = \frac{F_{\nu} d^2}{B_{\nu}(T_{d,c}) \kappa_\nu}$, where $F_{\nu}$ is the measured flux at 1.33 mm, $d$ is the distance to the source, $B_\nu$ is the Planck function at the corresponding dust temperature $T_d,c$, and $\kappa_\nu$ is the mass absorption coefficient that we take as $\kappa_\nu$ = 2 cm$^2$ g$^{-1}$ following \cite{Nilsson10}. In order to estimate a temperature for the dust, we have fitted a modified black-body to the available photometry at wavelengths longer than 10 $\mu m$ (AKARI, WISE, IRAS, Herschel)\footnote{Obtained from the CDS portal http://cdsportal.u-strasbg.fr} and the new ALMA photometry using the \textit{emcee} Affine Invariant Markov chain Monte Carlo Ensemble sampler implementation \citep{emcee}. This wavelength range was chosen to avoid the silicate emission observed in the mid-IR spectra reported by \cite{lisse17} around $\sim$10~$\mu m$. Additionally, there is some discrepancy between the photometric data of WISE, Herschel, and ALMA, and those of IRAS and AKARI. Given their larger PSFs and lower sensitivites, we decided to exclude the latter two from the fitting. The modified black body model assumes all dust grains have the same composition and size, and  accounts for changes in the dust emission efficiency ($Q_\lambda$) via two additional free parameters: $\beta$ and a reference wavelength $\lambda_0$, such that the black body emission is modified by a factor $Q_\lambda = 1 - \exp[(\lambda_0/\lambda)^\beta]$ \citep[e.g.][]{williams2004} Therefore, our model has four free parameters: a scaling factor which controls the disk luminosity, the dust temperature, $\beta$ and $\lambda_0$. Uniform priors were used for all parameters within reasonable ranges: scaling values that produce disk luminosities consistent with the observed photometry, dust temperatures between 20 and 250\,K, $\beta$ values between 0 and 1.5, and $\lambda_0$ between 0.3 and 300\,$\mu$m (the latter was explored in log scale). This analysis yields a dust temperature value of 153 $\pm$ 3 K, and a $\beta$ value of 0.24 $^{+0.07}_{-0.05}$ as shown in Fig. \ref{fig:fit}). $\lambda_0$ is unconstrained. Adopting this temperature value results in a dust mass of (9.0 $\pm$ 1.0) $\cdot$ 10$^{-2}$ M$_{\earth}$. Additionally, the resulting models yield $L_{\rm disk}/L*=(4.43 \pm 0.15)\times10^{-3}$, compatible with previous results (see Table \ref{tab:sample}).

\subsubsection{Gas}
\label{sect:gas}
We detect both $^{12}$CO (J=2-1, 230.538 GHz) and $^{13}$CO (J=2-1, 220.399 GHz), but we do not detect the C$^{18}$O (J=2-1, 219.560\,GHz) transition at 219 GHz (see upper panels of Fig. \ref{fig:HD36546}). Table \ref{tab:hd36546} lists the measured flux and peak intensity, and the size and angle of both lines. We estimate the  optical depth of the gas (see \cite{Lyo11}) as:
\begin{equation}
    \frac{F_{^{12}CO}}{F_{^{13}CO}} = \left(\frac{\nu _{^{12}CO}}{\nu _{^{13}CO}}\right)^{2} \times \frac{1-e^{-\tau_{12}}}{1-e^{-\tau_{12}/X}} ,
\end{equation}

where $X$ is the $^{12}$CO/$^{13}$CO abundance ratio \citep[$\sim$ 77, ][]{wilson94}. Following this expression we obtain $\tau_{12_{CO}} \sim$ 5, and $\tau_{13_{CO}} \sim$ 0.06, meaning $^{12}$CO and $^{13}$CO are optically thick and optically thin respectively.

Following \cite{moor17} (section 4), we calculated the CO mass as $M_{CO}={4 \pi m d^2}\frac{S_{21}}{x_{2}h\nu_{21}A_{21}} f_{iso}$; where $m$ is the mass of the CO molecule, $d$ is the distance to the star and $h$ is the Planck constant. We used the $^{13}$CO(J=2-1) as optically thin line for the flux $S_{21}$, frequency $\nu_{21}$ and Einstein constant $A_{21}$.  $f_{iso}$ is the isotope ratio ${{^{12}C ^{16}O}/ {^{13}C ^{16}O}}$, that was obtained from \cite{wilson94}, and $x_2$ is the fractional population of the upper level, calculated from the Boltzmann distribution, and considering a temperature of the gas equal to the excitation temperature of the CO ($\sim$ 20 K). This estimation is valid only under the assumption of LTE, with a high density gas and frequent collisions. In a non-LTE situation, with a lower density of the gas and the excitation of the molecules being dominated by radiative processes, the excitation temperature could be lower and therefore the gas mass would be higher \citep[for a more detailed explanation, see Sect. 4 in ][]{moor17}. Under this assumption, we obtain a CO mass of (3.2$\pm$1.2)$\cdot$10$^{-3}$M$_{\earth}$.

\begin{figure}
    \centering
	\includegraphics[width=.5\textwidth]{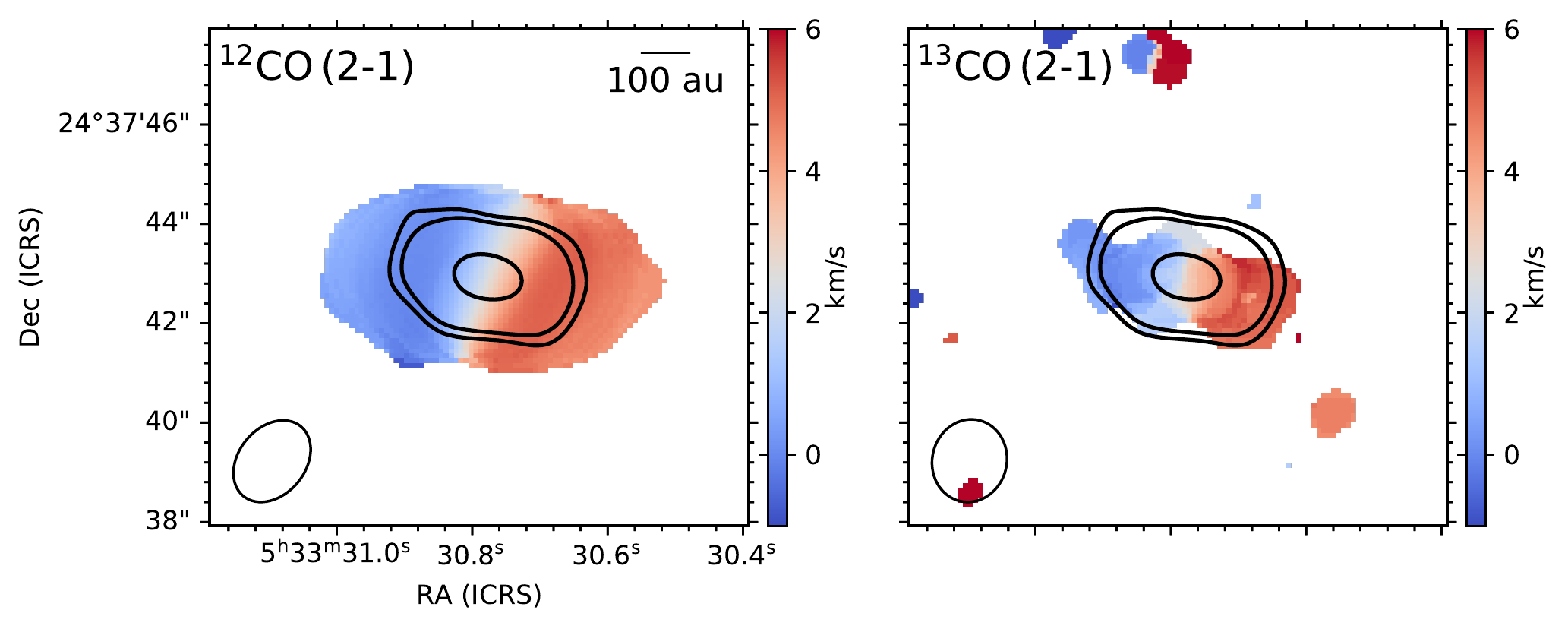}
    \caption{Velocity map of HD 36546 for $^{12}$CO and $^{13}$CO lines. Size of the beam is indicated in the lower left side. Contours show 5, 10 and 50 $\sigma$ of the continuum emission detections.}
    \label{fig:mom1_HD36546}
\end{figure}

The Keplerian rotation of the disc can be seen both in the spectral line profiles and in the velocity maps. Fig. \ref{fig:mom1_HD36546} displays the velocity maps for both lines, colour-coded to show the rotation of the disc. Bottom panels of Fig. \ref{fig:HD36546} show the spectral profile of the lines, where the double-peak emission is visible correspondent to the blue- and red-shifted regions of the gaseous disc with Keplerian rotation (note that the $^{13}$CO spectrum is binned with a factor of two to increase the S/N in each channel). The center of the line shows the systemic velocity of the object in the local kinematic standard of rest $v_{\rm helio} = 14.7\pm0.6$\,km~s$^{-1}$ that corresponds to a velocity of 3.3 $\pm$0.6\,$\mathrm{km s^{-1}}$ in the LSR frame \citep[see][]{Rebollido20}.

\begin{figure*}
    \centering
	\includegraphics[width=.9\textwidth]{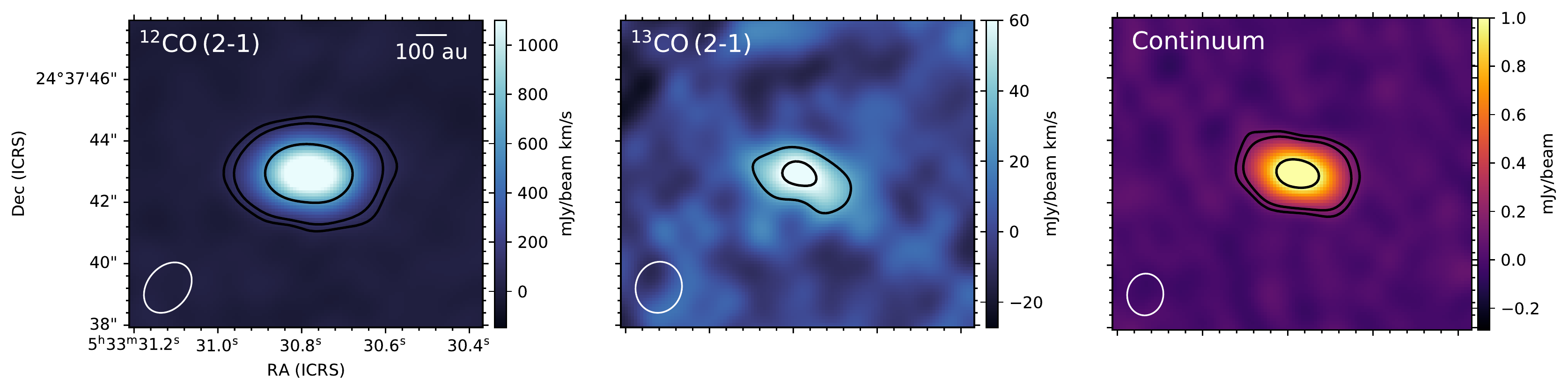}
	\includegraphics[width=.3\textwidth]{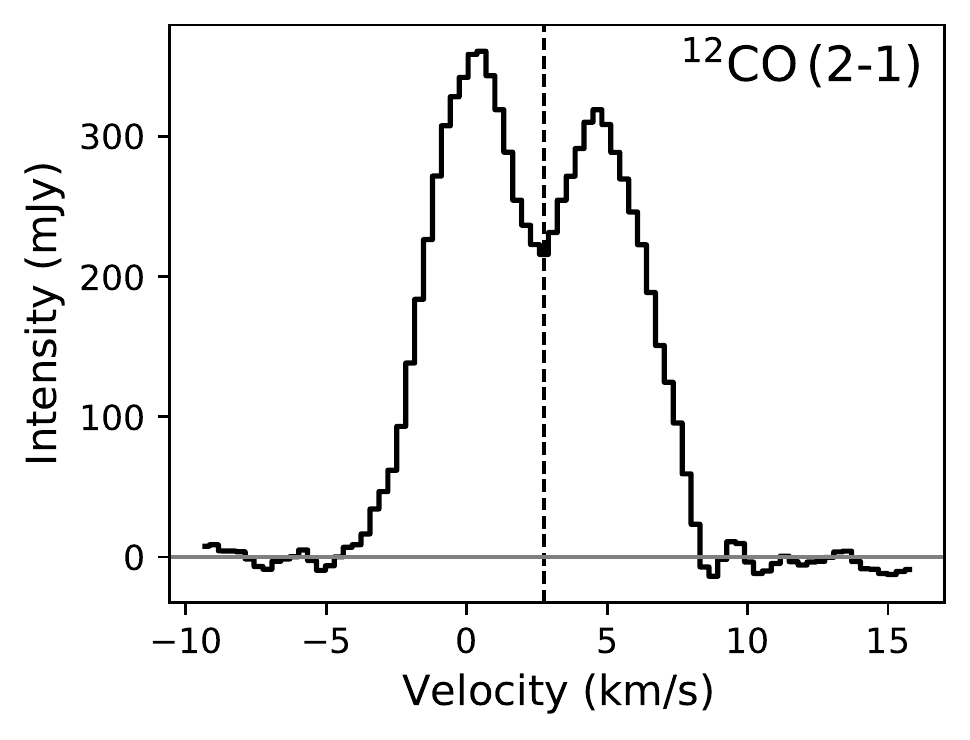}
	\includegraphics[width=.3\textwidth]{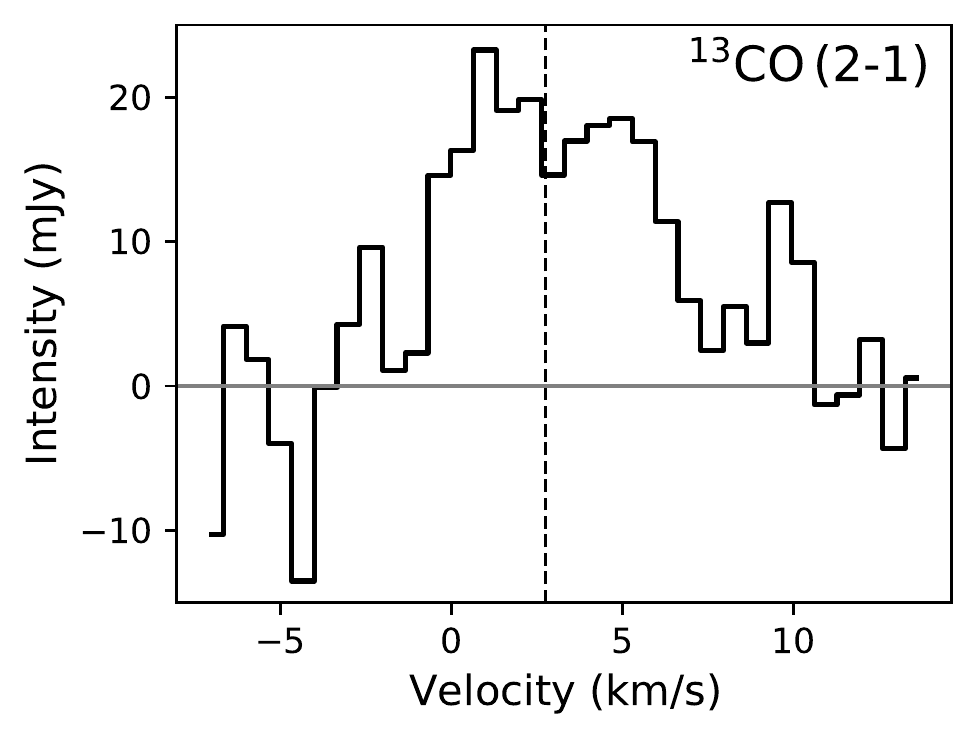}
    \caption{Emission observed in HD 36546. Top panels, left to right: Integrated emission of $^{12}$CO (230.538 GHz), $^{13}$CO (220.398 GHz) and continuum (1.33 mm). In all panels the size of the beam is indicated in the lower left corner. Contours indicate 5, 10 and 50 $\sigma$. Bottom panel, left to right: Spectra of $^{12}$CO and $^{13}$CO. The $^{12}$CO panel shows the double peaked profile, typically indicative of Keplerian rotation, and less visible in the $^{13}$CO panel. The continuous horizontal line marks the continuum (at 0 mJy) and the vertical dashed line shows the systemic velocity of the disc. The $^{13}$CO spectrum has been binned by a factor of two to improve the S/N.}
    \label{fig:HD36546}
\end{figure*}

\section{Discussion} 

\subsection{Cold gas in debris discs}

The initial goal of the observations was to determine if the presence of hot gas found in optical wavelengths and located in the inner regions of some debris disc systems could serve as a proxy for the presence of cold gas, much further out, and detected in long wavelengths (far-IR or sub-mm) as suggested in \cite{Rebollido18}. 
All objects in the observed sample show gaseous absorptions that can be interpreted as gas located in the inner regions of the system, very close to the star \citep[see Sect. \ref{sect:sample} and][]{Rebollido20}. 
We found cold gas in only one out of 8 sources, meaning there are no detections of cold gas in 87\% of the sample. This means at this level of sensitivity, we cannot rely on hot gas as an indicator for cold gas in debris discs, and deeper studies are needed.

\begin{figure*}
    \centering
	\includegraphics[width=1\textwidth]{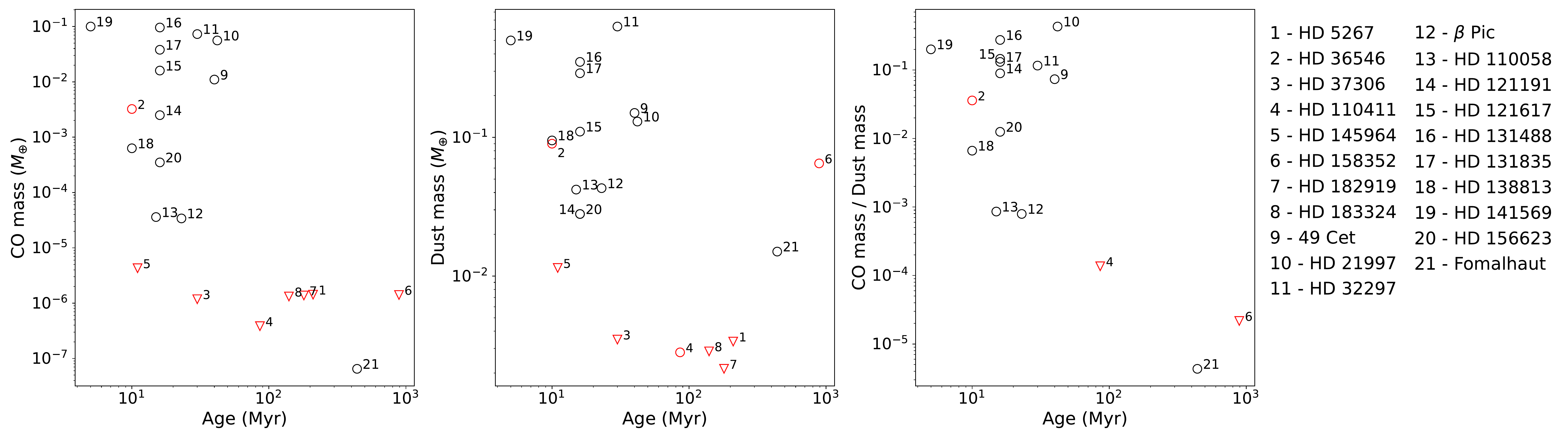}
    \caption{Context of the sample compared to other debris discs around A-type stars with detected CO gas. Red symbols mark the positions of the stars in the sample, numbered from 1 to 8. Triangles indicate upper-limits for non-detections. Black circles mark other debris discs with detections in the literature numbered from 9 to 21 \citep{moor17,moor19,DiFolco20}.}
    \label{fig:masses}
\end{figure*}

Fig. \ref{fig:masses} shows the gas and dust masses, and the gas to dust ratio of the objects in the sample, the debris disc with available CO gas masses around A-type stars \citep[CO and dust masses from][]{moor17,moor19}, and the hybrid disc candidate HD 141569 \citep{DiFolco20}. For the objects in our sample with non-detections, we used the upper limits in Table \ref{tab:sample} to roughly estimate the dust and CO mass limits following the procedures in Sect. \ref{sect:dust} and \ref{sect:gas}. The presence of CO gas cannot be discarded for the objects without detections, but the mass upper limits are very low in comparison with other gas bearing debris discs (left panel of Fig. \ref{fig:masses}).

According to the model of secondary gas discs \cite{kral17b}, the CO mass depends both on the CO production rate and the photodissociation lifetime of the released molecules which is influenced by the strength of the UV radiation field and the possible shielding processes (self-shielding of CO and shielding by C atoms). 
Assuming that dust and gas stems from an ongoing collisional cascade of icy planetesimals, the CO production rate can be estimated as M$_{\rm CO}$ = M$_{\rm loss}$ $\times$ $\gamma$, where $\gamma$ is the CO+CO$_2$ ice mass fraction of the solids, while $M_{\rm loss}$ is the mass loss of the cascade which has a strong dependence on the fractional luminosity ($\propto$ (L${\rm_{IR}}$/L$_*$)$^2$). 
Most currently known CO-bearing debris discs around A-type stars typically have fractional luminosities $>$5$\times$10$^{-4}$ which is much higher than the values typical of our sample (with the exception of HD 36546, our only detection) suggesting significantly larger CO production rates.  
Fomalhaut is the only one among them whose fractional luminosity is similar to that of our targets. \cite{matra17b} reported an integrated CO (2-1) line flux of 68 mJy\,kms$^{-1}$ for this object. Considering that even our closest target (HD 110411) is at least 5 times as far away as Fomalhaut, we could not have detected such a faint CO disc in our project. 
Therefore, the non-detections here do not necessarily rule out the presence of CO gas in our discs and thus are not enough to state that there is no correlation between the presence of hot and cold gas.

While we expect most of the gas to be of secondary origin in debris discs \citep{kral19}, the majority of objects with detected cold gas are younger than 100 Myr and it has been suggested that in young discs at least part of the gas could be primordial, i. e., hybrid discs \citep{kospal13}. 

A part of the observed sample is significantly older than most of the objects with detected gas, with the exception of HD 36546, HD 145964 and HD 37306. Still, Fomalhaut \citep[440 Myr]{matra17b} and $\eta$ Crv \citep[1 Gyr]{marino17} seem to host a non-negligible amount of gas, most likely of secondary origin given their large ages.

From Fig. \ref{fig:masses} (right panel), there is also no trend of the gas to dust ratio with age. The largest ratio is found in one of the oldest stars, HD 21997 (40 Myr), while the lowest is in $\beta$ Pic, a 20 Myr old star. Most of the objects clump over a gas to dust ratio of $\sim$ 3$\cdot$10$^{-2}$, but that is only 8 out of 13 (60\%), and they span ages from 3 Myr (lowest age estimation for HD 36546) to 42 Myr.
The sample might not be large enough to discuss statistics (only 14 discs with a reported CO mass, including HD 141569 and 15 with reported dust mass), and in order to investigate the temporal evolution of the gaseous component, we might need to differentiate between hybrid discs and secondary discs.

\subsection{The context of HD 36546}
Despite having a remarkably bright and close debris disc ($L_{\mathrm IR}/L_*$ = 3.4$\times$10$^{-3}$, see Table \ref{tab:sample}), there are few studies centred on HD 36546. This source is classified in the literature as a B8-A1 type star \citep[e.g.][]{Rebollido20,lisse17}, located at a distance of $\sim$ 100 pc \citep[Gaia EDR3,][]{GaiaDR3}. \cite{currie17} pointed out that the star could be linked to a sub-group of the Taurus-Auriga forming region, and estimated an age of 3-10 Myr. This would agree with the conclusion reached in \citep{lisse17}, that this object is in fact a very young debris disc that has recently lost his primary gaseous component. 

As many other debris disc \citep[e.g., see ][]{chen14}, HD 36546 seems to host a two-belt disc, of 570 and 135 K, as reported by \cite{lisse17} when performing a black body fit to both spectroscopic and photometric points. 
The outer region of the disc would be located further than 20 au, and corresponds to that imaged by \cite{currie17}, while the inner disc remains undetected in scattered light. 
A recent publication by \cite{lawson2021}, reporting multiwavelength scattered light observations in the near-IR, provides a new morphological analysis of the disc, consistent with previous results, including this paper.
Unfortunately, due to the original goal of the observations presented here (gas detection) the experimental design lacks the spatial resolution to detect any substructures in the disc. However, and while the sizes of the disc are similar (Table \ref{tab:hd36546}), the gas disc is slightly larger in the ALMA observations, as seen in Fig. \ref{fig:mom1_HD36546}. In that same figure the inclination of the gaseous disc seems to be the same as the dusty disc, in agreement with the reported values in Table \ref{tab:hd36546}. However, and due to the low spatial resolution, it is not possible to reach a firm conclusion.

Spectra in the mid-IR region shows an unusual feature at 7-8~$\mu$m along with the typical silicate feature around 10~$\mu$m, which could be indicative of a large abundance of carbon-rich mineral species. A recent giant collision is proposed as the source of this material, and in this scenario a large amount of gas would also have been released \citep{lisse17}.
According to \cite{fernandez06}, an overabundance of ionized carbon gas in the disc could act as a self-shield and prevent the blowout of the gas via Coulomb interactions \citep[as seen for other discs, e.g.][]{roberge00}, allowing the detection of hot stable gas such as the one reported by \cite{Rebollido20} for this object. The possibility of a dynamically active environment is also reinforced by the detection of exocomet-like events \citep[also reported in ][]{Rebollido20}, which have been suggested to require a planetary mass perturber \citep{beust91}. \cite{currie17} investigated the possible presence of a planet in the scattered light images that could affect the stirring of the disc (see Fig. 4 in their work), reaching upper limits of 5-6 M$_{J}$ at 23 au, and 2.5 M$_{J}$ at wider separations (>70 au). 

Finally, the difference in the detected disc size reported here for the continuum ($\sim$ 187 au) with respect to scattered light in near-IR \citep[170 au,][]{currie17,lawson2021} is not really significant, and might indicate a uniform grain size distribution in the disc. The orientation of the disc reported here is also compatible with the one reported in \cite{currie17,lawson2021}. Observations with a higher spatial resolution will allow a more precise determination of the distribution of the largest (mm) grains. 

\section{Summary}

The original goal of the observations presented here was to determine if there is a correlation between the presence of hot optical gas, and cold (sub-)mm gas, complementing the work in \cite{Rebollido18}. From a sample of eight debris discs, all of them with optical gas detected, we only detect (sub-)mm gas in one case, HD 36546. Even if the upper limits of the gas mass for these objects are very sensitive, this is not necessarily an indication of the lack of gas in these systems. Models such as the one presented in \citep{kral17b} suggest that the mass of CO is well below the detection limits of ALMA and APEX for discs with low IR luminosities, and therefore more sensitive observations would be required in order to reach a reliable conclusion. 
We also report continuum detection for three objects in the sample: HD 36546, HD 110411 and HD 158352.

We report here for the first time the detection of $^{12}$CO, $^{13}$CO and continuum at 1.3 mm in the debris disc of HD 36546. This object might be one of the youngest debris discs, with a reported age between 3 to 10 Myr. The detected gas content is similar to the rest of the known discs with gas, compatible with the model presented in \cite{kral17b}. While the origin of the gas in the disc is not clear, and could be a similar case to other objects present in the literature \citep[see e.g.][]{kospal13, pericaud17} where the origin of the gas is mostly primordial, or even a mix of both primordial and secondary gas, recent studies suggest that a large gas to dust ratio in debris discs can be explained with a secondary origin alone \citep{kral19}.
The context of this object, an unusually young bright debris disc ($L_{IR}/L_{*} \sim 10^{-3}$) with a carbon rich disc, the presence of exocomets detected in spectroscopy, and an age compatible with the end of its transition phase, makes it a promising candidate for follow up studies that could be key to unveil planet-disc interaction processes.

\section*{Acknowledgements}
This paper makes use of the following ALMA data: ADS/JAO.ALMA 2019.1.01517.S. ALMA is a partnership of ESO (representing its member states), NSF (USA) and NINS (Japan), together with NRC (Canada), MOST and ASIAA (Taiwan), and KASI (Republic of Korea), in cooperation with the Republic of Chile. The Joint ALMA Observatory is operated by ESO, AUI/NRAO and NAOJ. The National Radio Astronomy Observatory is a facility of the National Science Foundation operated under cooperative agreement by Associated Universities, Inc.
This work has made use of data from the European Space Agency (ESA) mission {\it Gaia} (\url{https://www.cosmos.esa.int/gaia}), processed by the {\it Gaia} Data Processing and Analysis Consortium (DPAC, \url{https://www.cosmos.esa.int/web/gaia/dpac/consortium}). Funding for the DPAC has been provided by national institutions, in particular the institutions participating in the {\it Gaia} Multilateral Agreement.
I.R., B.M. and E.V. were supported by the Spanish grant PGC2018-101950-B-I00 and the “María  de  Maeztu” award to Centro de Astrobiología (MDM-2017-0737). A.M acknowledges the support of the Hungarian National Research, Development  and  Innovation  Office  NKFIH  Grant  KH-130526. IdG-M is partially supported by MCIU-AEI (Spain) grant AYA2017-84390-C2-R (co-funded by FEDER).

\section*{Data Availability}

The data used in this paper (project 2019.1.01517.S) is publicly available in the ALMA archive (https://almascience.nrao.edu/asax/).



\bibliographystyle{mnras}
\bibliography{hd36546}

\begin{thebibliography}{}
\makeatletter
\relax
\def\mn@urlcharsother{\let\do\@makeother \do\$\do\&\do\#\do\^\do\_\do\%\do\~}
\def\mn@doi{\begingroup\mn@urlcharsother \@ifnextchar [ {\mn@doi@}
  {\mn@doi@[]}}
\def\mn@doi@[#1]#2{\def\@tempa{#1}\ifx\@tempa\@empty \href
  {http://dx.doi.org/#2} {doi:#2}\else \href {http://dx.doi.org/#2} {#1}\fi
  \endgroup}
\def\mn@eprint#1#2{\mn@eprint@#1:#2::\@nil}
\def\mn@eprint@arXiv#1{\href {http://arxiv.org/abs/#1} {{\tt arXiv:#1}}}
\def\mn@eprint@dblp#1{\href {http://dblp.uni-trier.de/rec/bibtex/#1.xml}
  {dblp:#1}}
\def\mn@eprint@#1:#2:#3:#4\@nil{\def\@tempa {#1}\def\@tempb {#2}\def\@tempc
  {#3}\ifx \@tempc \@empty \let \@tempc \@tempb \let \@tempb \@tempa \fi \ifx
  \@tempb \@empty \def\@tempb {arXiv}\fi \@ifundefined
  {mn@eprint@\@tempb}{\@tempb:\@tempc}{\expandafter \expandafter \csname
  mn@eprint@\@tempb\endcsname \expandafter{\@tempc}}}

\bibitem[\protect\citeauthoryear{{Abrahamyan}, {Mickaelian}  \&
  {Knyazyan}}{{Abrahamyan} et~al.}{2015}]{Abrahamyan15}
{Abrahamyan} H.~V.,  {Mickaelian} A.~M.,   {Knyazyan} A.~V.,  2015, \mn@doi
  [Astronomy and Computing] {10.1016/j.ascom.2014.12.002}, \href
  {https://ui.adsabs.harvard.edu/abs/2015A&C....10...99A} {10, 99}

\bibitem[\protect\citeauthoryear{{Bell}, {Mamajek}  \& {Naylor}}{{Bell}
  et~al.}{2015}]{bell15}
{Bell} C.~P.~M.,  {Mamajek} E.~E.,   {Naylor} T.,  2015, \mn@doi [\mnras]
  {10.1093/mnras/stv1981}, \href
  {http://adsabs.harvard.edu/abs/2015MNRAS.454..593B} {454, 593}

\bibitem[\protect\citeauthoryear{{Beust}, {Vidal-Madjar}  \& {Ferlet}}{{Beust}
  et~al.}{1991}]{beust91}
{Beust} H.,  {Vidal-Madjar} A.,   {Ferlet} R.,  1991, \aap, \href
  {http://adsabs.harvard.edu/abs/1991A%26A...247..505B} {247, 505}

\bibitem[\protect\citeauthoryear{{Chen}, {Mittal}, {Kuchner}, {Forrest},
  {Lisse}, {Manoj}, {Sargent}  \& {Watson}}{{Chen} et~al.}{2014}]{chen14}
{Chen} C.~H.,  {Mittal} T.,  {Kuchner} M.,  {Forrest} W.~J.,  {Lisse} C.~M.,
  {Manoj} P.,  {Sargent} B.~A.,   {Watson} D.~M.,  2014, \mn@doi [\apjs]
  {10.1088/0067-0049/211/2/25}, \href
  {http://cdsads.u-strasbg.fr/abs/2014ApJS..211...25C} {211, 25}

\bibitem[\protect\citeauthoryear{{Currie} et~al.,}{{Currie}
  et~al.}{2017}]{currie17}
{Currie} T.,  et~al., 2017, \mn@doi [\apjl] {10.3847/2041-8213/836/1/L15},
  \href {http://adsabs.harvard.edu/abs/2017ApJ...836L..15C} {836, L15}

\bibitem[\protect\citeauthoryear{{Cutri} et~al.,}{{Cutri}
  et~al.}{2013}]{cutri13}
{Cutri} R.~M.,  et~al., 2013, {Explanatory Supplement to the AllWISE Data
  Release Products}, Explanatory Supplement to the AllWISE Data Release
  Products

\bibitem[\protect\citeauthoryear{{Di Folco}, {P{\'e}ricaud}, {Dutrey},
  {Augereau}, {Chapillon}, {Guilloteau}, {Pi{\'e}tu}  \& {Boccaletti}}{{Di
  Folco} et~al.}{2020}]{DiFolco20}
{Di Folco} E.,  {P{\'e}ricaud} J.,  {Dutrey} A.,  {Augereau} J.~C.,
  {Chapillon} E.,  {Guilloteau} S.,  {Pi{\'e}tu} V.,   {Boccaletti} A.,  2020,
  \mn@doi [\aap] {10.1051/0004-6361/201732243}, \href
  {https://ui.adsabs.harvard.edu/abs/2020A&A...635A..94D} {635, A94}

\bibitem[\protect\citeauthoryear{{Draper}, {Matthews}, {Kennedy}, {Wyatt},
  {Venn}  \& {Sibthorpe}}{{Draper} et~al.}{2016}]{draper16}
{Draper} Z.~H.,  {Matthews} B.~C.,  {Kennedy} G.~M.,  {Wyatt} M.~C.,  {Venn}
  K.~A.,   {Sibthorpe} B.,  2016, \mn@doi [\mnras] {10.1093/mnras/stv2696},
  \href {http://adsabs.harvard.edu/abs/2016MNRAS.456..459D} {456, 459}

\bibitem[\protect\citeauthoryear{ESA}{ESA}{1997}]{ESA1997}
ESA ed. 1997, {The HIPPARCOS and TYCHO catalogues. Astrometric and photometric
  star catalogues derived from the ESA HIPPARCOS Space Astrometry Mission}  ESA
  Special Publication Vol. 1200

\bibitem[\protect\citeauthoryear{{Egan} \& {Price}}{{Egan} \&
  {Price}}{1996}]{Egan96}
{Egan} M.~P.,  {Price} S.~D.,  1996, \mn@doi [\aj] {10.1086/118227}, \href
  {https://ui.adsabs.harvard.edu/abs/1996AJ....112.2862E} {112, 2862}

\bibitem[\protect\citeauthoryear{{Ferlet}, {Vidal-Madjar}  \& {Hobbs}}{{Ferlet}
  et~al.}{1987}]{ferlet87}
{Ferlet} R.,  {Vidal-Madjar} A.,   {Hobbs} L.~M.,  1987, \aap, \href
  {http://adsabs.harvard.edu/abs/1987A%26A...185..267F} {185, 267}

\bibitem[\protect\citeauthoryear{{Fern{\'a}ndez}, {Brandeker}  \&
  {Wu}}{{Fern{\'a}ndez} et~al.}{2006}]{fernandez06}
{Fern{\'a}ndez} R.,  {Brandeker} A.,   {Wu} Y.,  2006, \mn@doi [\apj]
  {10.1086/500788}, \href {http://adsabs.harvard.edu/abs/2006ApJ...643..509F}
  {643, 509}

\bibitem[\protect\citeauthoryear{{Foreman-Mackey}, {Hogg}, {Lang}  \&
  {Goodman}}{{Foreman-Mackey} et~al.}{2013}]{emcee}
{Foreman-Mackey} D.,  {Hogg} D.~W.,  {Lang} D.,   {Goodman} J.,  2013, \mn@doi
  [\pasp] {10.1086/670067}, \href
  {https://ui.adsabs.harvard.edu/abs/2013PASP..125..306F} {125, 306}

\bibitem[\protect\citeauthoryear{{Gaia Collaboration} et~al.,}{{Gaia
  Collaboration} et~al.}{2018}]{GAIADR2}
{Gaia Collaboration} et~al., 2018, \mn@doi [\aap]
  {10.1051/0004-6361/201833051}, \href
  {https://ui.adsabs.harvard.edu/abs/2018A&A...616A...1G} {616, A1}

\bibitem[\protect\citeauthoryear{{Gaia Collaboration}, {Brown}, {Vallenari},
  {Prusti}, {de Bruijne}, {Babusiaux}  \& {Biermann}}{{Gaia Collaboration}
  et~al.}{2020}]{GaiaDR3}
{Gaia Collaboration} {Brown} A.~G.~A.,  {Vallenari} A.,  {Prusti} T.,  {de
  Bruijne} J.~H.~J.,  {Babusiaux} C.,   {Biermann} M.,  2020, arXiv e-prints,
  \href {https://ui.adsabs.harvard.edu/abs/2020arXiv201201533G} {p.
  arXiv:2012.01533}

\bibitem[\protect\citeauthoryear{{Hobbs}, {Vidal-Madjar}, {Ferlet}, {Albert}
  \& {Gry}}{{Hobbs} et~al.}{1985}]{hobbs85}
{Hobbs} L.~M.,  {Vidal-Madjar} A.,  {Ferlet} R.,  {Albert} C.~E.,   {Gry} C.,
  1985, \mn@doi [\apjl] {10.1086/184485}, \href
  {http://adsabs.harvard.edu/abs/1985ApJ...293L..29H} {293, L29}

\bibitem[\protect\citeauthoryear{{Hobbs}, {Welty}, {Lagrange-Henri}, {Ferlet}
  \& {Vidal-Madjar}}{{Hobbs} et~al.}{1988}]{hobbs88}
{Hobbs} L.~M.,  {Welty} D.~E.,  {Lagrange-Henri} A.~M.,  {Ferlet} R.,
  {Vidal-Madjar} A.,  1988, \mn@doi [\apjl] {10.1086/185308}, \href
  {http://adsabs.harvard.edu/abs/1988ApJ...334L..41H} {334, L41}

\bibitem[\protect\citeauthoryear{{Hughes}, {Wilner}, {Kamp}  \&
  {Hogerheijde}}{{Hughes} et~al.}{2008}]{hughes08}
{Hughes} A.~M.,  {Wilner} D.~J.,  {Kamp} I.,   {Hogerheijde} M.~R.,  2008,
  \mn@doi [\apj] {10.1086/588520}, \href
  {http://adsabs.harvard.edu/abs/2008ApJ...681..626H} {681, 626}

\bibitem[\protect\citeauthoryear{{Hughes}, {Duch{\^e}ne}  \&
  {Matthews}}{{Hughes} et~al.}{2018}]{hughes18}
{Hughes} A.~M.,  {Duch{\^e}ne} G.,   {Matthews} B.~C.,  2018, \mn@doi [\araa]
  {10.1146/annurev-astro-081817-052035}, \href
  {https://ui.adsabs.harvard.edu/abs/2018ARA&A..56..541H} {56, 541}

\bibitem[\protect\citeauthoryear{{Kiefer}, {Lecavelier des Etangs}, {Boissier},
  {Vidal-Madjar}, {Beust}, {Lagrange}, {H{\'e}brard}  \& {Ferlet}}{{Kiefer}
  et~al.}{2014}]{kiefer14b}
{Kiefer} F.,  {Lecavelier des Etangs} A.,  {Boissier} J.,  {Vidal-Madjar} A.,
  {Beust} H.,  {Lagrange} A.-M.,  {H{\'e}brard} G.,   {Ferlet} R.,  2014,
  \mn@doi [\nat] {10.1038/nature13849}, \href
  {http://adsabs.harvard.edu/abs/2014Natur.514..462K} {514, 462}

\bibitem[\protect\citeauthoryear{{K{\'o}sp{\'a}l} et~al.,}{{K{\'o}sp{\'a}l}
  et~al.}{2013}]{kospal13}
{K{\'o}sp{\'a}l} {\'A}.,  et~al., 2013, \mn@doi [\apj]
  {10.1088/0004-637X/776/2/77}, \href
  {http://adsabs.harvard.edu/abs/2013ApJ...776...77K} {776, 77}

\bibitem[\protect\citeauthoryear{{Kral}, {Matr{\`a}}, {Wyatt}  \&
  {Kennedy}}{{Kral} et~al.}{2017}]{kral17b}
{Kral} Q.,  {Matr{\`a}} L.,  {Wyatt} M.~C.,   {Kennedy} G.~M.,  2017, \mn@doi
  [\mnras] {10.1093/mnras/stx730}, \href
  {https://ui.adsabs.harvard.edu/abs/2017MNRAS.469..521K} {469, 521}

\bibitem[\protect\citeauthoryear{{Kral}, {Marino}, {Wyatt}, {Kama}  \&
  {Matr{\`a}}}{{Kral} et~al.}{2019}]{kral19}
{Kral} Q.,  {Marino} S.,  {Wyatt} M.~C.,  {Kama} M.,   {Matr{\`a}} L.,  2019,
  \mn@doi [\mnras] {10.1093/mnras/sty2923}, \href
  {https://ui.adsabs.harvard.edu/abs/2019MNRAS.489.3670K} {489, 3670}

\bibitem[\protect\citeauthoryear{{Kurucz}}{{Kurucz}}{1993}]{kurucz93}
{Kurucz} R.,  1993, SYNTHE Spectrum Synthesis Programs and Line Data.~Kurucz
  CD-ROM No.~18.~Cambridge, Mass.: Smithsonian Astrophysical Observatory,
  1993., \href {http://adsabs.harvard.edu/abs/1993KurCD..18.....K} {18}

\bibitem[\protect\citeauthoryear{{Lawson} et~al.,}{{Lawson}
  et~al.}{2021}]{lawson2021}
{Lawson} K.,  et~al., 2021, arXiv e-prints, \href
  {https://ui.adsabs.harvard.edu/abs/2021arXiv210908984L} {p. arXiv:2109.08984}

\bibitem[\protect\citeauthoryear{{Lieman-Sifry}, {Hughes}, {Carpenter},
  {Gorti}, {Hales}  \& {Flaherty}}{{Lieman-Sifry} et~al.}{2016}]{liemansifry16}
{Lieman-Sifry} J.,  {Hughes} A.~M.,  {Carpenter} J.~M.,  {Gorti} U.,  {Hales}
  A.,   {Flaherty} K.~M.,  2016, \mn@doi [\apj] {10.3847/0004-637X/828/1/25},
  \href {http://adsabs.harvard.edu/abs/2016ApJ...828...25L} {828, 25}

\bibitem[\protect\citeauthoryear{{Lisse}, {Sitko}, {Russell}, {Marengo},
  {Currie}, {Melis}, {Mittal}  \& {Song}}{{Lisse} et~al.}{2017}]{lisse17}
{Lisse} C.~M.,  {Sitko} M.~L.,  {Russell} R.~W.,  {Marengo} M.,  {Currie} T.,
  {Melis} C.,  {Mittal} T.,   {Song} I.,  2017, \mn@doi [\apjl]
  {10.3847/2041-8213/aa6ea3}, \href
  {http://adsabs.harvard.edu/abs/2017ApJ...840L..20L} {840, L20}

\bibitem[\protect\citeauthoryear{{Lyo}, {Ohashi}, {Qi}, {Wilner}  \&
  {Su}}{{Lyo} et~al.}{2011}]{Lyo11}
{Lyo} A.~R.,  {Ohashi} N.,  {Qi} C.,  {Wilner} D.~J.,   {Su} Y.-N.,  2011,
  \mn@doi [\aj] {10.1088/0004-6256/142/5/151}, \href
  {https://ui.adsabs.harvard.edu/abs/2011AJ....142..151L} {142, 151}

\bibitem[\protect\citeauthoryear{{Marino} et~al.,}{{Marino}
  et~al.}{2016}]{marino16}
{Marino} S.,  et~al., 2016, \mn@doi [\mnras] {10.1093/mnras/stw1216}, \href
  {http://adsabs.harvard.edu/abs/2016MNRAS.460.2933M} {460, 2933}

\bibitem[\protect\citeauthoryear{{Marino} et~al.,}{{Marino}
  et~al.}{2017}]{marino17}
{Marino} S.,  et~al., 2017, \mn@doi [\mnras] {10.1093/mnras/stw2867}, \href
  {http://adsabs.harvard.edu/abs/2017MNRAS.465.2595M} {465, 2595}

\bibitem[\protect\citeauthoryear{{Marton} et~al.,}{{Marton}
  et~al.}{2017}]{HPSC17}
{Marton} G.,  et~al., 2017, arXiv e-prints, \href
  {https://ui.adsabs.harvard.edu/abs/2017arXiv170505693M} {p. arXiv:1705.05693}

\bibitem[\protect\citeauthoryear{{Matr{\`a}} et~al.,}{{Matr{\`a}}
  et~al.}{2017a}]{matra17a}
{Matr{\`a}} L.,  et~al., 2017a, \mn@doi [\mnras] {10.1093/mnras/stw2415}, \href
  {http://adsabs.harvard.edu/abs/2017MNRAS.464.1415M} {464, 1415}

\bibitem[\protect\citeauthoryear{{Matr{\`a}} et~al.,}{{Matr{\`a}}
  et~al.}{2017b}]{matra17b}
{Matr{\`a}} L.,  et~al., 2017b, \mn@doi [\apj] {10.3847/1538-4357/aa71b4},
  \href {http://adsabs.harvard.edu/abs/2017ApJ...842....9M} {842, 9}

\bibitem[\protect\citeauthoryear{{McMullin}, {Waters}, {Schiebel}, {Young}  \&
  {Golap}}{{McMullin} et~al.}{2007}]{McMullin07}
{McMullin} J.~P.,  {Waters} B.,  {Schiebel} D.,  {Young} W.,   {Golap} K.,
  2007, in {Shaw} R.~A.,  {Hill} F.,   {Bell} D.~J.,  eds,  Astronomical
  Society of the Pacific Conference Series Vol. 376, Astronomical Data Analysis
  Software and Systems XVI. p.~127

\bibitem[\protect\citeauthoryear{{Mo{\'o}r} et~al.,}{{Mo{\'o}r}
  et~al.}{2011}]{moor11b}
{Mo{\'o}r} A.,  et~al., 2011, \mn@doi [\apjl] {10.1088/2041-8205/740/1/L7},
  \href {http://adsabs.harvard.edu/abs/2011ApJ...740L...7M} {740, L7}

\bibitem[\protect\citeauthoryear{{Mo{\'o}r} et~al.,}{{Mo{\'o}r}
  et~al.}{2015}]{moor15a}
{Mo{\'o}r} A.,  et~al., 2015, \mn@doi [\apj] {10.1088/0004-637X/814/1/42},
  \href {http://adsabs.harvard.edu/abs/2015ApJ...814...42M} {814, 42}

\bibitem[\protect\citeauthoryear{{Mo{\'o}r} et~al.,}{{Mo{\'o}r}
  et~al.}{2017}]{moor17}
{Mo{\'o}r} A.,  et~al., 2017, \mn@doi [\apj] {10.3847/1538-4357/aa8e4e}, \href
  {http://adsabs.harvard.edu/abs/2017ApJ...849..123M} {849, 123}

\bibitem[\protect\citeauthoryear{{Mo{\'o}r} et~al.,}{{Mo{\'o}r}
  et~al.}{2019}]{moor19}
{Mo{\'o}r} A.,  et~al., 2019, \mn@doi [\apj] {10.3847/1538-4357/ab4272}, \href
  {https://ui.adsabs.harvard.edu/abs/2019ApJ...884..108M} {884, 108}

\bibitem[\protect\citeauthoryear{{Nilsson} et~al.,}{{Nilsson}
  et~al.}{2010}]{Nilsson10}
{Nilsson} R.,  et~al., 2010, \mn@doi [\aap] {10.1051/0004-6361/201014444},
  \href {https://ui.adsabs.harvard.edu/abs/2010A&A...518A..40N} {518, A40}

\bibitem[\protect\citeauthoryear{{P{\'e}ricaud}, {Di Folco}, {Dutrey},
  {Guilloteau}  \& {Pi{\'e}tu}}{{P{\'e}ricaud} et~al.}{2017}]{pericaud17}
{P{\'e}ricaud} J.,  {Di Folco} E.,  {Dutrey} A.,  {Guilloteau} S.,
  {Pi{\'e}tu} V.,  2017, \mn@doi [\aap] {10.1051/0004-6361/201629371}, \href
  {https://ui.adsabs.harvard.edu/abs/2017A&A...600A..62P} {600, A62}

\bibitem[\protect\citeauthoryear{{Pilbratt} et~al.,}{{Pilbratt}
  et~al.}{2010}]{pilbratt10}
{Pilbratt} G.~L.,  et~al., 2010, \mn@doi [\aap] {10.1051/0004-6361/201014759},
  \href {https://ui.adsabs.harvard.edu/abs/2010A&A...518L...1P} {518, L1}

\bibitem[\protect\citeauthoryear{{Rebollido} et~al.,}{{Rebollido}
  et~al.}{2018}]{Rebollido18}
{Rebollido} I.,  et~al., 2018, \mn@doi [\aap] {10.1051/0004-6361/201732329},
  \href {http://cdsads.u-strasbg.fr/abs/2018A%26A...614A...3R} {614, A3}

\bibitem[\protect\citeauthoryear{{Rebollido} et~al.,}{{Rebollido}
  et~al.}{2020}]{Rebollido20}
{Rebollido} I.,  et~al., 2020, \mn@doi [\aap] {10.1051/0004-6361/201936071},
  \href {https://ui.adsabs.harvard.edu/abs/2020A&A...639A..11R} {639, A11}

\bibitem[\protect\citeauthoryear{{Riviere-Marichalar}
  et~al.,}{{Riviere-Marichalar} et~al.}{2014}]{riviere14}
{Riviere-Marichalar} P.,  et~al., 2014, \mn@doi [\aap]
  {10.1051/0004-6361/201322901}, \href
  {http://adsabs.harvard.edu/abs/2014A%26A...565A..68R} {565, A68}

\bibitem[\protect\citeauthoryear{{Roberge} \& {Weinberger}}{{Roberge} \&
  {Weinberger}}{2008}]{roberge08}
{Roberge} A.,  {Weinberger} A.~J.,  2008, \mn@doi [\apj] {10.1086/527314},
  \href {http://adsabs.harvard.edu/abs/2008ApJ...676..509R} {676, 509}

\bibitem[\protect\citeauthoryear{{Roberge}, {Feldman}, {Lagrange},
  {Vidal-Madjar}, {Ferlet}, {Jolly}, {Lemaire}  \& {Rostas}}{{Roberge}
  et~al.}{2000}]{roberge00}
{Roberge} A.,  {Feldman} P.~D.,  {Lagrange} A.~M.,  {Vidal-Madjar} A.,
  {Ferlet} R.,  {Jolly} A.,  {Lemaire} J.~L.,   {Rostas} F.,  2000, \mn@doi
  [\apj] {10.1086/309157}, \href
  {http://adsabs.harvard.edu/abs/2000ApJ...538..904R} {538, 904}

\bibitem[\protect\citeauthoryear{{Roberge} et~al.,}{{Roberge}
  et~al.}{2013}]{roberge13}
{Roberge} A.,  et~al., 2013, \mn@doi [\apj] {10.1088/0004-637X/771/1/69}, \href
  {http://adsabs.harvard.edu/abs/2013ApJ...771...69R} {771, 69}

\bibitem[\protect\citeauthoryear{{Thureau} et~al.,}{{Thureau}
  et~al.}{2014}]{thureau14}
{Thureau} N.~D.,  et~al., 2014, \mn@doi [\mnras] {10.1093/mnras/stu1864}, \href
  {http://adsabs.harvard.edu/abs/2014MNRAS.445.2558T} {445, 2558}

\bibitem[\protect\citeauthoryear{{Williams}, {Najita}, {Liu}, {Bottinelli},
  {Carpenter}, {Hillenbrand}, {Meyer}  \& {Soderblom}}{{Williams}
  et~al.}{2004}]{williams2004}
{Williams} J.~P.,  {Najita} J.,  {Liu} M.~C.,  {Bottinelli} S.,  {Carpenter}
  J.~M.,  {Hillenbrand} L.~A.,  {Meyer} M.~R.,   {Soderblom} D.~R.,  2004,
  \mn@doi [\apj] {10.1086/381721}, \href
  {https://ui.adsabs.harvard.edu/abs/2004ApJ...604..414W} {604, 414}

\bibitem[\protect\citeauthoryear{{Wilson} \& {Rood}}{{Wilson} \&
  {Rood}}{1994}]{wilson94}
{Wilson} T.~L.,  {Rood} R.,  1994, \mn@doi [\araa]
  {10.1146/annurev.aa.32.090194.001203}, \href
  {https://ui.adsabs.harvard.edu/abs/1994ARA&A..32..191W} {32, 191}

\bibitem[\protect\citeauthoryear{{Wyatt}}{{Wyatt}}{2018}]{wyatt18}
{Wyatt} M.~C.,  2018, {Debris Disks: Probing Planet Formation}.
p.~146, \mn@doi{10.1007/978-3-319-55333-7\_146}

\bibitem[\protect\citeauthoryear{{Zuckerman}, {Forveille}  \&
  {Kastner}}{{Zuckerman} et~al.}{1995}]{zuckerman95}
{Zuckerman} B.,  {Forveille} T.,   {Kastner} J.~H.,  1995, \mn@doi [\nat]
  {10.1038/373494a0}, \href {http://adsabs.harvard.edu/abs/1995Natur.373..494Z}
  {373, 494}

\makeatother
\end{thebibliography}




\newpage

\appendix

\section{Figures}

\begin{figure*}
    \centering
	\includegraphics[width=0.33\textwidth]{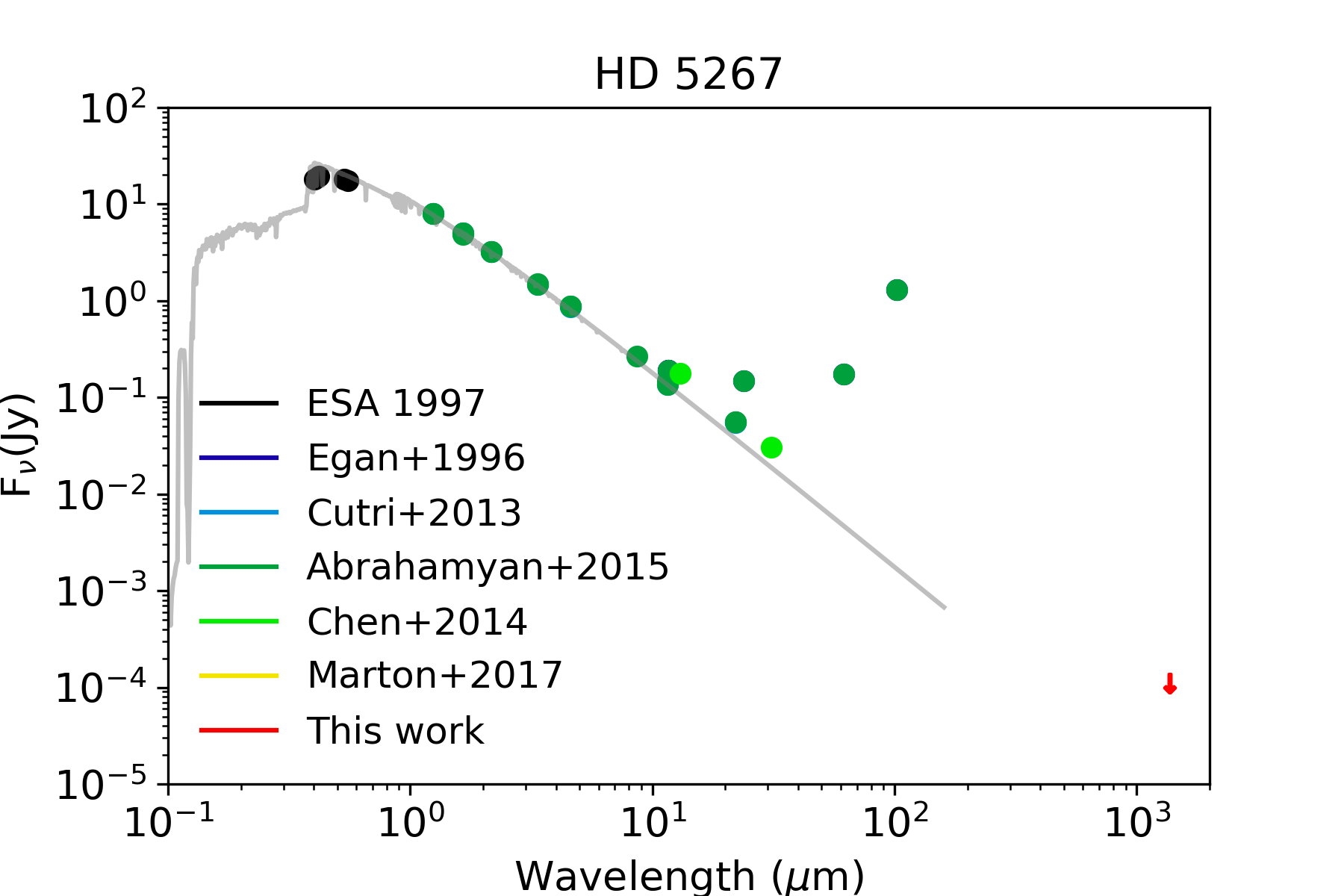}
	\includegraphics[width=0.33\textwidth]{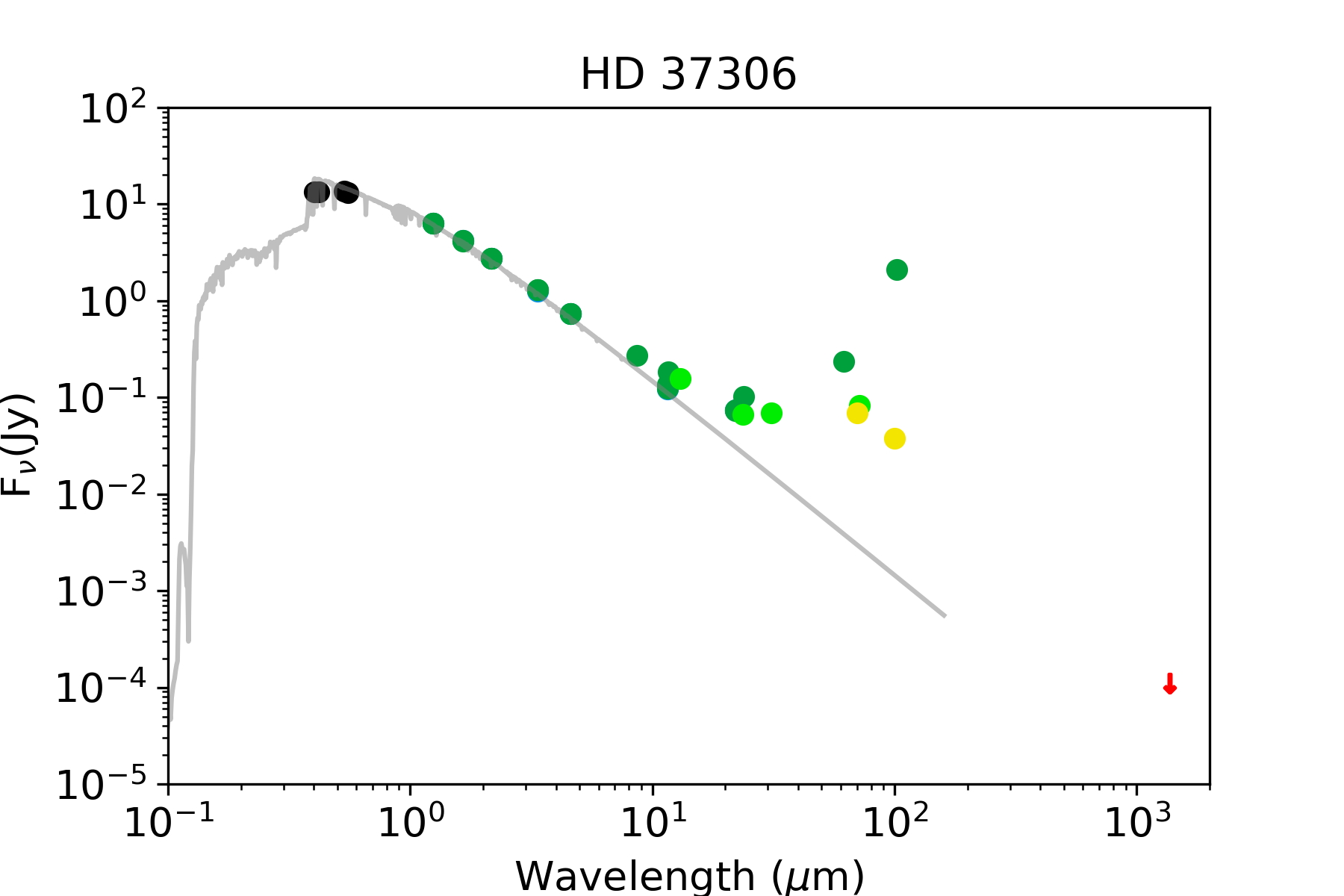}
	\includegraphics[width=0.33\textwidth]{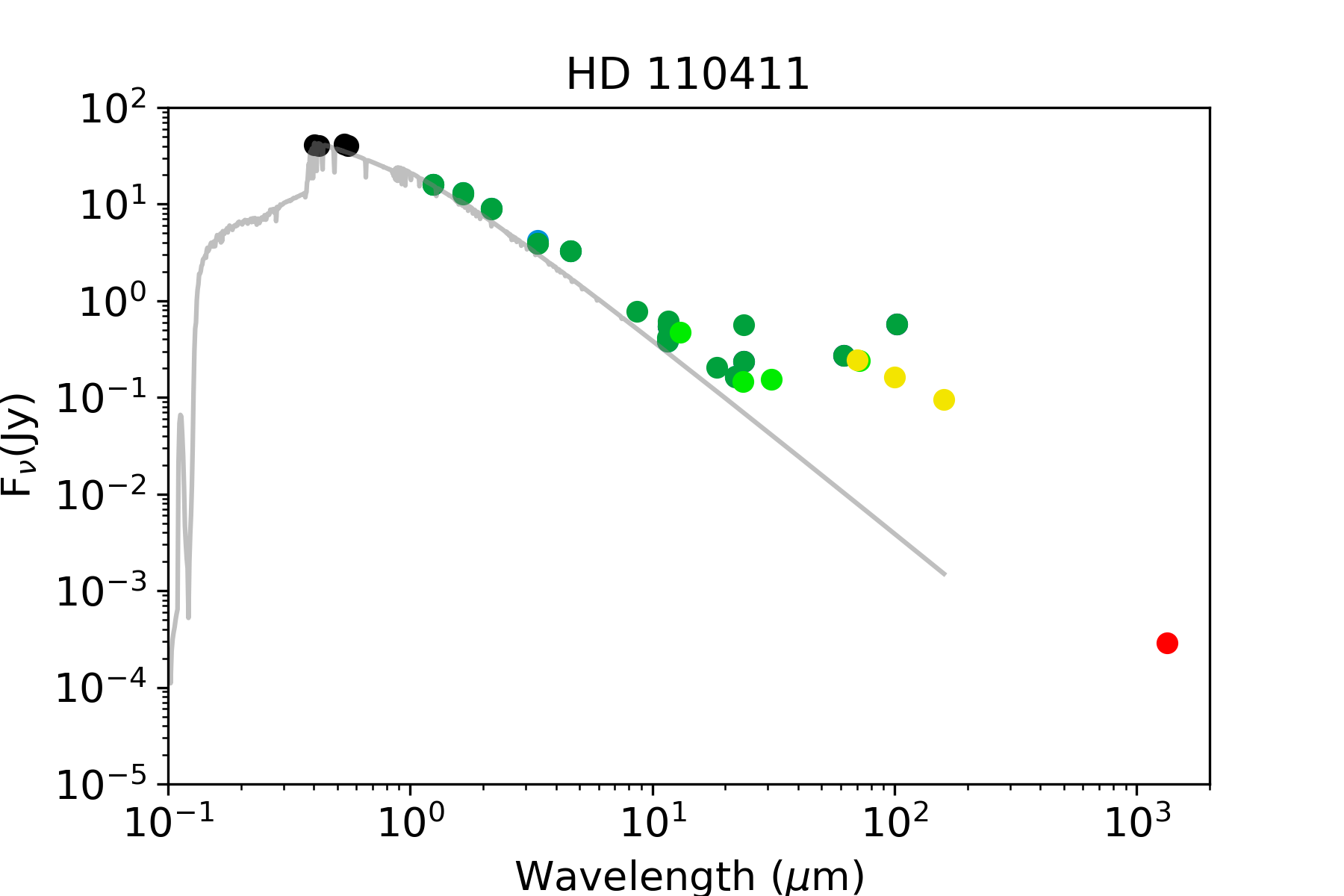}
	\includegraphics[width=0.33\textwidth]{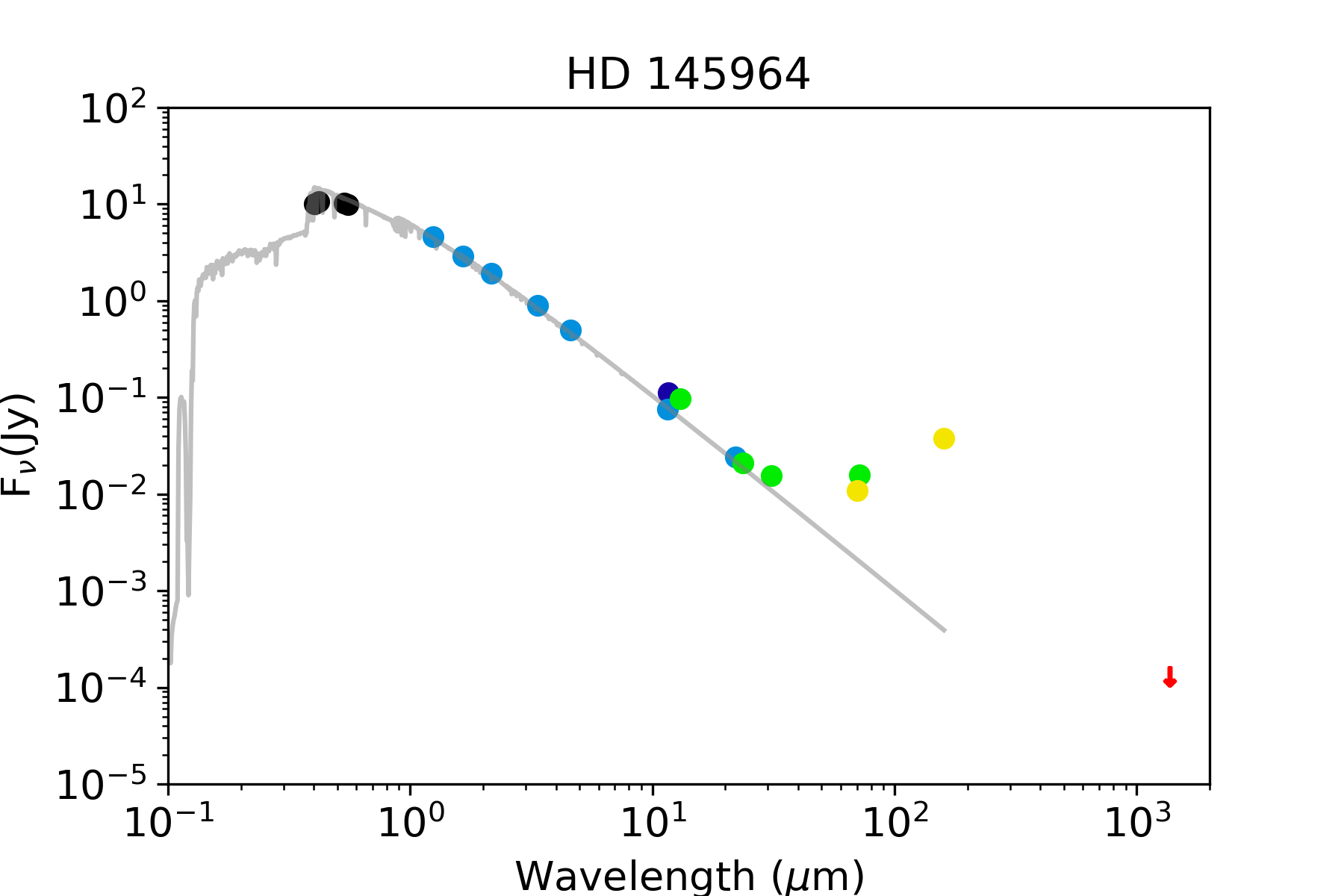}
	\includegraphics[width=0.33\textwidth]{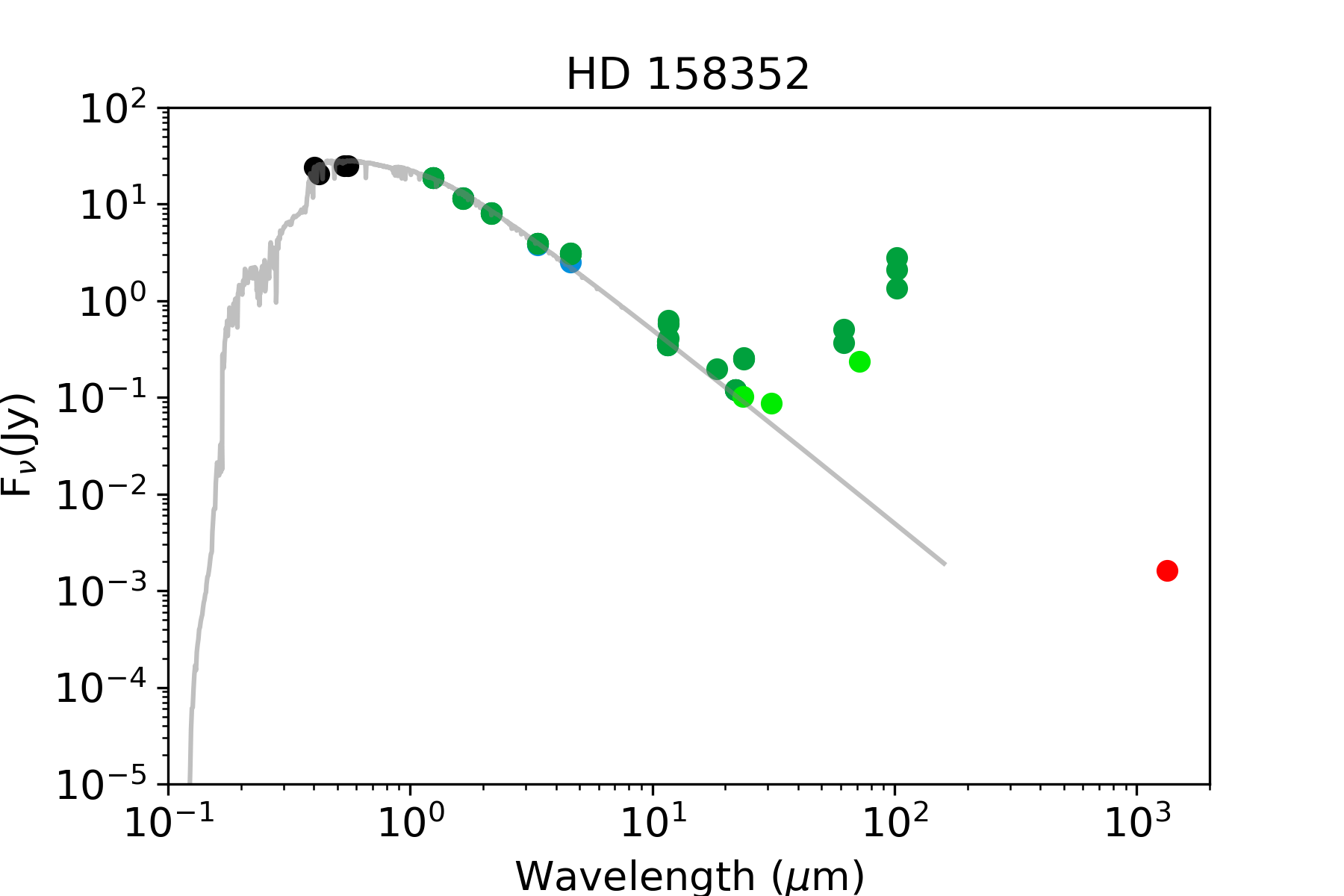}
	\includegraphics[width=0.33\textwidth]{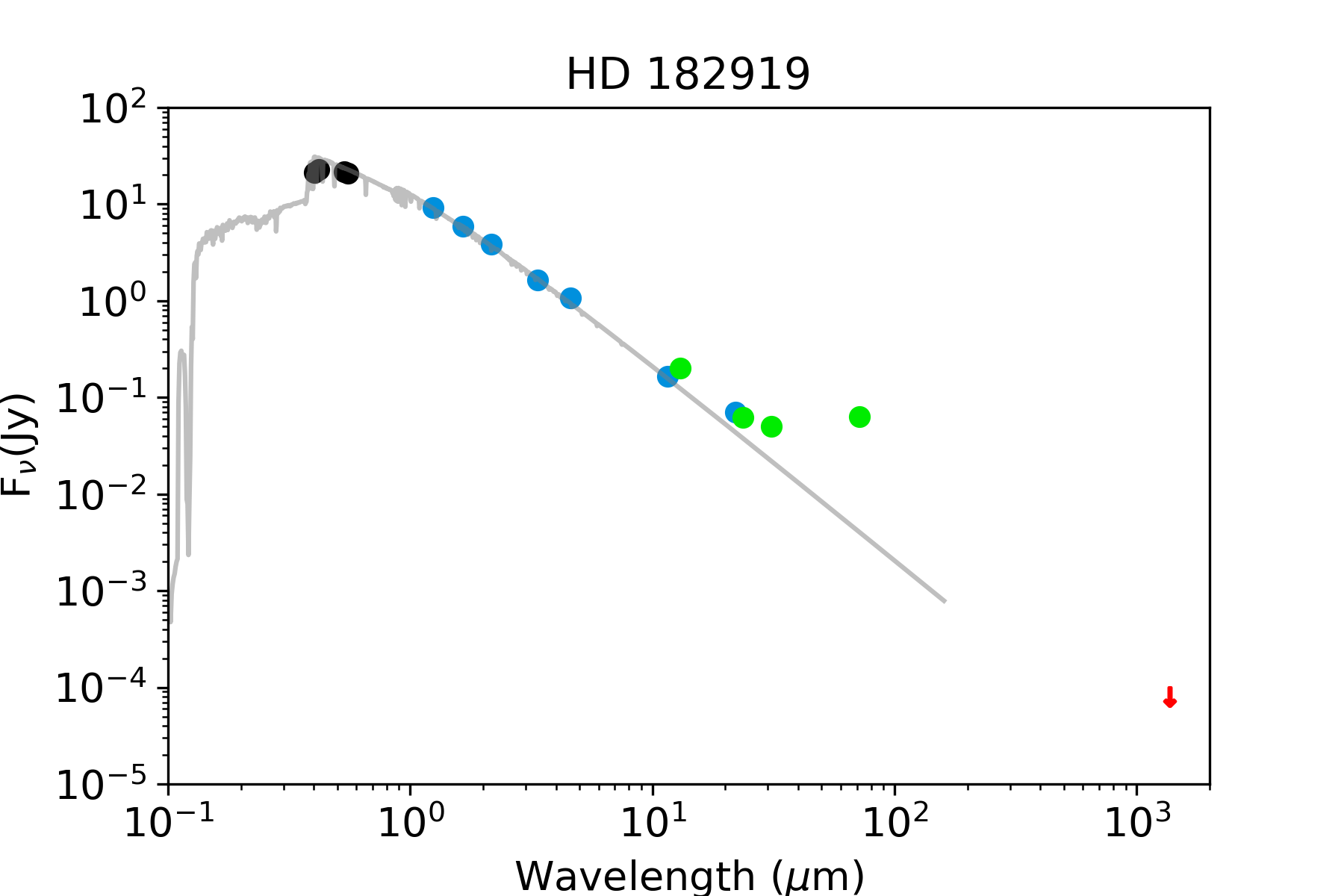}x
	\includegraphics[width=0.33\textwidth]{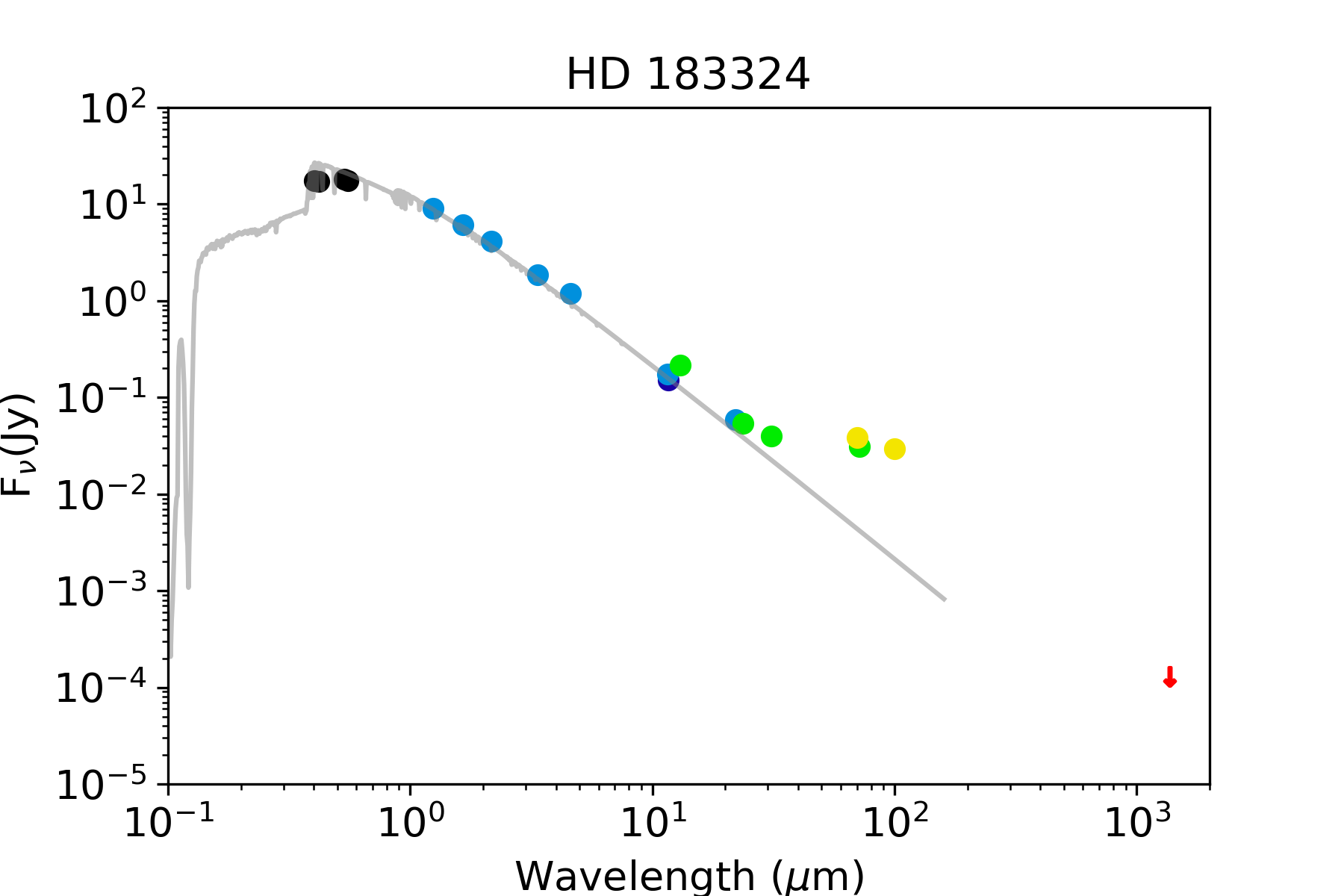}
	
    \caption{Spectral energy distribution of objects in the sample with no CO detections. Spectral models shown in gray were interpolated from \protect\cite{kurucz93} according to the paramters listed in \protect\cite{Rebollido20}. Arrows in ALMA fluxes indicate upper limits. References:\protect\cite{ESA1997,Egan96,cutri13,Abrahamyan15,chen14,HPSC17} }
    \label{fig:seds}
\end{figure*}

\begin{figure*}
    \centering
	\includegraphics[width=1.\textwidth]{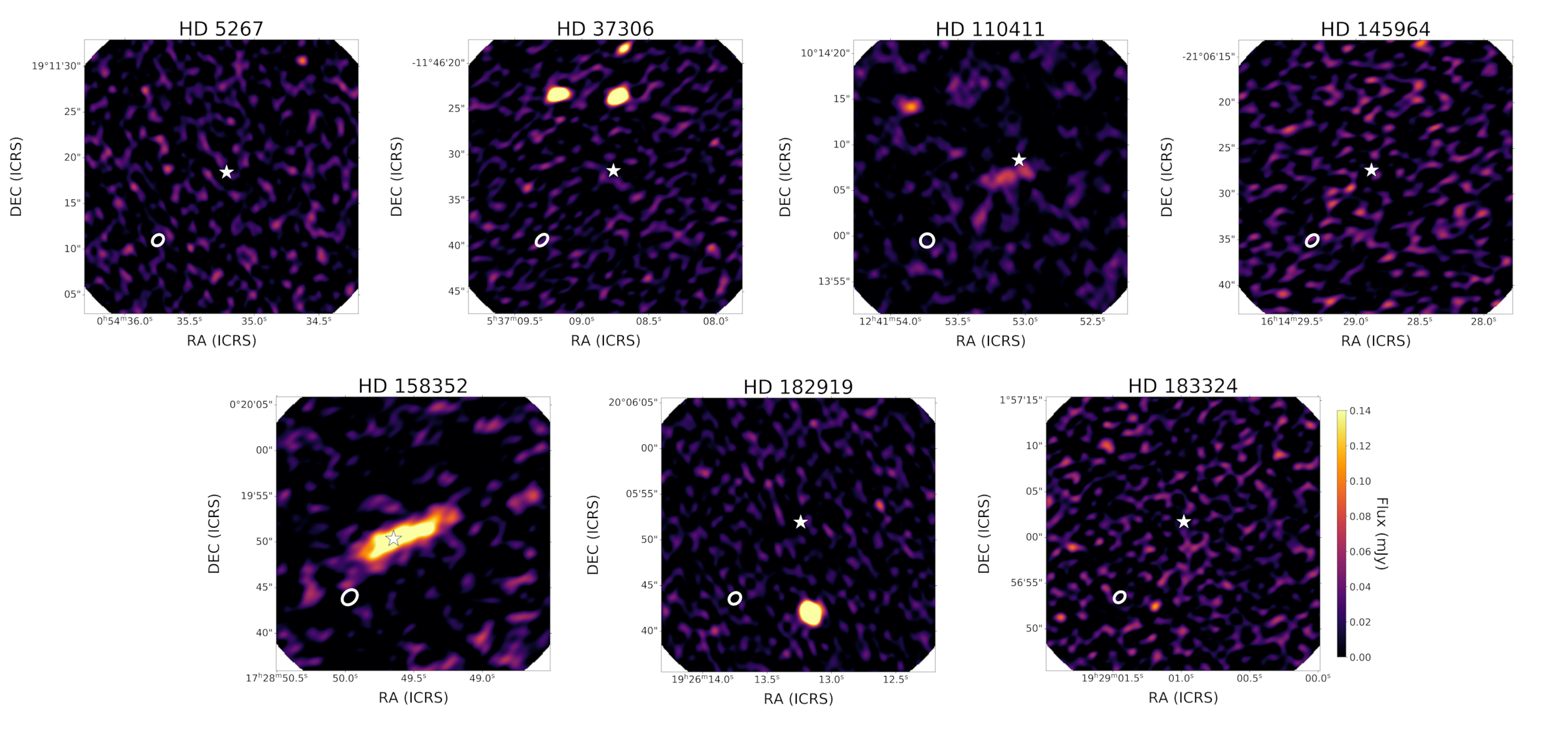}
    \caption{Maps of continuum emission at 1.3 mm for the objects in the sample with non-detected CO. ALMA beam size is shown in the bottom left of all figures. The nominal position of the stars is marked with a white star symbol. Four objects show continuum emission, two of them (HD 37306 and HD 181929) not compatible with the source location. See Sect. 3.1.}
    \label{fig:sample}
\end{figure*}



\bsp	
\label{lastpage}
\end{document}